\documentclass[hidelinks,twoside,11pt]{article}

\usepackage{amsmath}
\usepackage{wrapfig}
\usepackage{lipsum}
\usepackage{graphicx, xcolor, varwidth}
\usepackage{amsfonts}

\usepackage{caption}
\usepackage{subcaption}
\usepackage{pdflscape}
\usepackage[counterclockwise, figuresright]{rotating}
\usepackage[margin=1in]{geometry}
\usepackage{setspace}
\usepackage{hyperref}
\date{}

\usepackage[sort&compress, numbers]{natbib}

\usepackage{authblk}
\title{\vspace{-0.2in}\textbf{\Larger Dynamic Control of Stochastic Evolution: A Deep Reinforcement Learning Approach to Adaptively Targeting Emergent Drug Resistance}}
\author{\vspace{0.1in}Dalit Engelhardt\thanks{dengelhardt@ds.dfci.harvard.edu}}

\newcommand*\samethanks[1][\value{footnote}]{\footnotemark[#1]}

\affil{\footnotesize{Department of Chemistry and Chemical Biology, Harvard University, Cambridge, MA, USA}\vspace{-0.15in}}
\affil{\footnotesize{Department of Data Science, Dana-Farber Cancer Institute, Boston, MA, USA}\thanks{Current affiliations.}\vspace{-0.15in}}
\affil{\footnotesize{Department of Biostatistics, Harvard T. H. Chan School of Public Health, Boston, MA, USA}\samethanks\vspace{-0.15in}}
\affil{\footnotesize{Department of Stem Cell and Regenerative Biology, Harvard University, Cambridge, MA, USA}\samethanks\vspace{-0.15in}}
\affil{\footnotesize{Center for Cancer Evolution, Dana-Farber Cancer Institute, Boston, MA, USA}\samethanks}
\date{}                   

\begin{document}

\long\def\/*#1*/{}

\title{\textbf{\Large \begin{singlespace}\vspace{-1in}
			Dynamic Control of Stochastic Evolution: A Deep Reinforcement Learning Approach to Adaptively Targeting Emergent Drug Resistance
\end{singlespace}}}

\maketitle

\begin{abstract}%
	
The challenge in controlling stochastic systems in which low-probability events can set the system on catastrophic trajectories is to develop a robust ability to respond to such events without significantly compromising the optimality of the baseline control policy. This paper presents CelluDose, a stochastic simulation-trained deep reinforcement learning adaptive feedback control prototype for automated precision drug dosing targeting stochastic and heterogeneous cell proliferation. Drug resistance can emerge from random and variable mutations in targeted cell populations; in the absence of an appropriate dosing policy, emergent resistant subpopulations can proliferate and lead to treatment failure. Dynamic feedback dosage control holds promise in combatting this phenomenon, but the application of traditional control approaches to such systems is fraught with challenges due to the complexity of cell dynamics, uncertainty in model parameters, and the need in medical applications for a robust controller that can be trusted to properly handle unexpected outcomes. Here, training on a sample biological scenario identified single-drug and combination therapy policies that exhibit a $100\%$ success rate at suppressing cell proliferation and responding to diverse system perturbations while establishing low-dose no-event baselines. These policies were found to be highly robust to variations in a key model parameter subject to significant uncertainty and unpredictable dynamical changes.
\end{abstract}

\section{Introduction}

Advances in experimental capabilities and high-throughput data analytics  in recent years have contributed to a significant rise in biological and biomedical model complexity. Driven by the anticipation that higher-complexity models will lead to increased predictive power, these efforts may ultimately lead to an improved ability to effectively control real biological systems via successful biomedical interventions. Yet as model complexity rises the problem of system control becomes increasingly less tractable, necessitating a tradeoff between the ability to develop controllers for systems of interest and the predictive accuracy of the system model~\citep{parker2001control}. Significant and variable stochasticity, nonlinearity in the dynamics, a potentially high-dimensional space of biological units, e.g. cell types, and parameter uncertainty all pose significant challenges to the application of traditional control methods to complex biological systems models. In order to bridge the gap between advances in biological and pharmacological modeling and the ability to engineer appropriate biomedical controllers, approaches that can effectively control such systems will likely be required in the near future.

The application of model-free reinforcement learning methods to continuous control tasks has seen significant advances in recent
years\footnote{See, e.g.,~\cite{recht2018tour,kiumarsi2018optimal} for recent surveys.} and holds promise for the control of systems for which mathematical optimization and control may be intractable, with potential applications in diverse fields that include robotics, mathematical finance, and healthcare. However, for reinforcement learning to become a standard tool in control engineering and related fields more work is needed in the successful application of reinforcement learning continuous control methods to highly stochastic and realistic environments. In particular, many applications require the safe adaptability and robustness of the controller to unexpected feedback. In learning-based control policies where such theoretical guarantees may not be available, a verified ability to learn policies capable of efficient generalization across parameter values not seen during training is crucial for safe implementation. Similarly important is the ability to discover, despite system stochasticity and random events, high-preference low-cost policies applicable when such perturbations do not occur.

The focus of this work is on the development of a deep reinforcement learning  dynamic feedback
control prototype, CelluDose, for precision dosing that
adaptively targets harmful cell populations of variable drug susceptibility and
resistance levels based on discrete-time feedback on the targeted cell population structure. The development of drug resistance in response to therapy is a major cause of treatment failure worldwide across a variety of diseases,
patient populations, and administered drugs. The emergence of resistance during treatment is a complex, multi-dimensional stochastic biological process whose control requires a safe and effective balance between minimal drug administration and the proper targeting of high-resistance cell types, which arise randomly and with highly variable rates.  Resistance to a drug
can be present prior to treatment or it can emerge during treatment~\citep{holohan2013cancer,brusselaers2011rising,munita2016mechanisms} through diverse mechanisms. It evolves dynamically and often non-uniformly: intercellular variability can lead to faster adaptation to environmental pressures~\citep{bodi2017phenotypic} and thus promote the rise of resistant subpopulations in a previously-susceptible population of cells. Tumor heterogeneity is now understood to be a major contributor to drug resistance in cancer~\citep{dexter1986tumor,swanton2012intratumor,holohan2013cancer,dagogo2018tumour}, and variability in resistance among bacterial cell populations has been tied to treatment failure~\citep{falagas2008heteroresistance,band2018carbapenem}. In bacterial evolution experiments, the emergence of multiple antibiotic-resistant cell lines has been observed~\citep{toprak2012evolutionary,suzuki2014prediction}. Such clonal diversity, even in a majority-susceptible cell population, can be the harbinger of drug resistance that ultimately leads to treatment
failure: improper treatments can exert a selective evolutionary pressure
that leads to the elimination of susceptible cell populations but
leaves more resistant cells largely unaffected and able to proliferate
due to reduced competition with susceptible cells~\citep{read2014antibiotic}. 

The emergence and potential rise to dominance of resistant strains are stochastic processes driven by the dynamic and complex interaction between the influences and interplay of natural selection, demographic stochasticity, and environmental fluctuations. A predictive understanding of the likely evolutionary trajectories toward drug resistance is key to developing treatments that can effectively target and suppress resistant cell populations, but a fully predictive understanding of such processes remains a challenge. Control of evolution, however, does not in principle require full predictability or determinism in evolution. In a closed-loop setting, system feedback can mitigate the reliance on precise trajectory knowledge, so long as this feedback can be obtained at reasonable time intervals, the uncertainty and stochasticity can be approximated in an informed manner, and the controller is sufficiently robust to changes in system behavior and parameter fluctuations. 

In heterogeneous cell environments prone to resistance evolution, the development of such a control policy must balance the need for properly and efficiently targeting all cell types that may either exist at the onset of treatment or that may spontaneously emerge during therapy, while maintaining sufficiently low dosing for toxicity minimization. Inappropriate dosing
can lead to the proliferation of resistant cell populations, but this
interplay is subject to significant levels of stochasticity as well as uncertainty in model parameters that must
be accounted for in the design of the control algorithm. The target
goal of CelluDose is to combat the emergence of drug resistance during
treatment by sensitively adjusting dosage and/or switching the administered
drug in response to observed feedback on changes in the cell population
structure while employing minimal overall drug dosage for toxicity
reduction. CelluDose combines a model-free deep reinforcement learning algorithm with model information in developing an adaptive dosing control policy for the elimination of harmful cell populations with heterogeneous drug responses and undergoing random, low-probability, yet potentially significant demographic changes. 

Beyond presenting this implementation, this work also aims to motivate the use of model-free reinforcement learning control
in the development of next-generation precision dosing controllers
by demonstrating that a robust and highly responsive adaptive control policy
can be obtained for a complex and stochastic biological system that
is intractable with traditional control methods. While the focus of this work is on a particular class of biological scenarios, it should be noted that the incorporation of certain training elements described below may facilitate and stabilize the training of robust continuous control policies in other stochastic environments experiencing low-probability significant events. To that end, a discussion of practices in this work and subsequent insights that may be usefully generalized to other stochastic control environments is included.

\section{Background}

\subsection{Reinforcement learning for continuous control}\label{subsec:RLcontrolInto}

The aim of a therapeutic agent targeting harmful cell populations
is to produce an effect that leads to the elimination of this
cell population while minimizing toxicity by avoiding the use of high doses. Given the possibility
of emergence of high-resistance cell populations, this task involves
a potentially complex tradeoff between acceptably low dosing levels,
the need to effectively eliminate the targeted cell populations, and
a preference for shorter treatment times to reduce patient morbidity
and \textendash{} in infectious diseases \textendash{} the possibility
of further contagion. As a result, the dosing problem can be cast
in somewhat different ways that depend on the particular disease conditions
and prevailing treatment preferences. For the purposes of this work, a \textit{successful}
treatment is defined as one in which all targeted harmful cell subpopulations
were eliminated by some acceptable maximal treatment time. Within
this measure of success \textit{optimality} is defined as lowest expected
cumulative dosing over the course of the treatment. A treatment
course is designated as a \textit{failure} if any targeted cells remain
after the maximal treatment time has been reached, with the extent of failure dependent on the number of remaining cells.

Model-free reinforcement learning (RL) approaches aim to learn an optimal
decision-making policy through an iterative process in which the policy
is gradually improved by sequential tuning of parameters either of
the policy itself or of a value (e.g. the action-value function) indicative
of the policy's optimality. The data on which learning is done is
supplied in the form of transitions from a previous state $s$ of
the environment to the next state $s'$ that occurs probabilistically
(in stochastic environments) when a certain action $a$ is chosen.
Each such transition is assigned a reward $r$, and reward maximization
drives the optimization of the policy. Learning can be done on a step-by-step
basis or episodically, where a single episode consists of multiple
steps of transitions. In the case of the dosing problem considered
here, an episode is a single-patient full course of therapy. It ends when either the maximal time has been reached
or all harmful cells have been eliminated, whichever comes first. Episodes are thus finite but can be long, with drugs
administered at discrete time steps over a continuous range of dosages.
At each time interval, a decision is made on which drugs at what doses
should be administered based on observations of the current state
of the disease, defined as the concentrations of the targeted cell
types. However, only at the end of each episode does it become clear
whether or not the sequence of therapeutic actions was in fact successful.
This credit assignment problem~\citep{minsky1961steps} frequently plagues episodic
tasks and will be addressed in Section~\ref{subsec:RL}.

The optimal policy for this decision-making process needs to provide
next-time-step dosing guidelines under the optimality guidelines described
above and based on potential stochastic evolutionary scenarios described
by the model in Section \ref{sec:biological-model}. This system is
continuous in its state space (cell population composition), involves
time-varying and potentially high stochasticity, can be high-dimensional
due to the large number of possibly-occurring mutant cell types, and
involves one or multiple controls (drugs) that if administered intravenously
(the scenario of interest here) can in principle take a continuous range of values. For reasons of mechanical
operability and medical explainability, a deterministic dosing
policy is needed in this case.

Deep deterministic policy gradient (DDPG)~\citep{lillicrap2015continuous}
is an off-policy actor-critic deterministic policy deep reinforcement learning
algorithm suitable for continuous and high-dimensional observation and
action spaces. Building on the deterministic policy gradient algorithm~\citep{silver2014deterministic},
DDPG employs neural network function approximation through the incorporation
of two improvements used in Deep Q Networks~\citep{mnih2013playing,mnih2015human} for increasing learning stability: the use
of target networks and a replay buffer from which mini-batches
of $(s,a,s',r)$ transitions are sampled during training. In the resistance evolution scenarios considered here, the observation
space (number of cell types) can be quite large, especially if finer
observations are available on cellular heterogeneity (finer differentiations
in dose responses), and it is continuous since cell density is treated here as a continuous variable in all but the lowest density levels. Although the action space is typically low-dimensional (only a few drugs are usually
considered for treatment of a particular disease), it is continuous in the intravenous administration case considered here. For these reasons, the dosing control scheme described here employs the DDPG algorithm. 

\subsection{Model-informed treatment planning with reinforcement learning}

An important benefit of a mechanistic model-informed approach to treatment planning is the ability to explore therapeutic decisions prior to the start of or as an accompaniment to clinical trials, which can then be used to inform clinical trials. The simulation-based use of RL here thus differs from work that employs reinforcement learning for making decisions based on patient data and clinical outcomes, where exploration of treatment parameter space is constrained by data available from previously-attempted treatments. When this data becomes available during and after clinical trials, these approaches and the mechanistic model-based approach presented here could be used in a complementary manner to inform and improve individualized treatments.

The use of RL for mechanistic model control in treatment planning is less common than data-driven approaches. We note below a few applications of interest. Q-learning~\citep{sutton2018reinforcement},
a discrete state and action space RL algorithm, was employed in~\cite{moore2014reinforcement}
for closed-loop control of propofol anesthesia and in~\cite{padmanabhan2017reinforcement} for control of cancer
chemotherapy based on an ordinary differential equations (ODE)-based
deterministic model of tumor growth. In~\cite{ahn2011drug} a natural
actor-critic algorithm~\citep{peters2008natural} was employed for
cancer chemotherapy control of a similar ODE-based model, with a binary
action space (drug or no drug). Tumors were taken to be uniform and resistance to treatment was not considered as part of the models used in~\cite{padmanabhan2017reinforcement} and in~\cite{ahn2011drug}. In~\cite{petersen2018precision} a DDPG-based algorithm was applied to an agent-based model of sepsis progression for obtaining an adaptive cytokine therapy policy.

 Interest in adaptive control of therapy that takes resistance evolution into account has been on the rise in recent years (see, e.g.,~\cite{fischer2015value,newton2019nonlinear}) but these efforts have been restricted to simple models that are typically low dimensional and/or deterministic. The approach presented here provides a mechanism for determining and automating dosing in a responsive and robust way that can be generalized to arbitrarily heterogeneous cell populations exhibiting complex and realistic dynamics. Both single-drug and combination therapy control policies are developed in this context, and observations on population composition are supplied at discrete time intervals, as would be expected in a laboratory or clinical scenario. By extension, this work seeks to motivate efforts in the high-resolution tracking of cell-level drug resistance progression within an individual patient. From a control engineering perspective, the incorporation of model information into model-free learning presented here may be usefully transported and adapted to the control of other stochastic systems with an equations-based description for improved RL learning stability and convergence; this and insights into robustness for learning-based control engineering are discussed in Sec.~\ref{sec:controlinsights}. For reinforcement learning insights, a detailed analysis is included of policy features that were observed in training.

\subsection{Implementation and workflow}

Training and implementation rely on three key components: model and simulation development, selection of the subset of drugs of interest, knowledge of the range of likely resistance levels that have been recorded in response to these drugs, and access to diagnostics that can provide temporal data on cell population structure. In this paper, all data supplied in training was simulated; a description of how such data may be obtained for future proof-of-concept validation is included below.

For demonstrative purposes and facility of near-future validation, the focus of this work is on drug resistance in evolving bacterial populations subjected to growth-inhibiting antibiotic drugs. We note that evolutionary models of cell population dynamics are also used to understand and model cancer progression and the development of resistance to treatment, and that another promising application area for a CelluDose-based platform is in cancer therapy control.

Future clinical implementation of a CelluDose-based controller could
be as an integrated system with the ability to track and supply the trained control model with sensitive measurements of the presence of small populations of cells and equipped
with an automated intravenous infusion mechanism. This implementation would be particularly useful in intensive care units (ICUs), where the development of resistance to therapy is a major concern, therapy is performed in the controlled environment of the clinic, and substantial genetic diversity in bacterial strains has been observed~\citep{roach2015year}. Alternatively, in longer-term therapies, it could be implemented
as a decision support tool providing informed dosing recommendations
to clinicians. In both cases, observations at the chosen fixed time intervals for which training was done would be supplied in the form of the respective targeted cell type concentrations, where a distinct cell type is (phenotypically) defined by non-negligible differences in the dose response across the possible ranges of dosages that are permitted for administration.

\subsubsection{Modeling and simulation}

Although training implements a model-free reinforcement learning algorithm, model knowledge is used in two ways: to provide the training data (as simulation data) and for algorithm design via feature engineering and reward assignment. This feature engineering is made possible by trajectory estimation from the equations-based component of the growth-mutation model implemented here. This model combines (1) a stochastic differential equations-based system for approximating the growth of existing cell subpopulations perturbed by (2) random events  -- mutations -- that create new subpopulations of resistant cells or increase such pre-existing populations. These events increase the effective dimensionality and alter the composition of the system at random points in time. The resulting system exhibits variable and substantial levels of stochasticity that can significantly alter its trajectory and potentially lead to treatment failure (see Fig.~\ref{fig:simsIis1}).  Since, as shown in Section~\ref{sec:results}, the learned policies are highly robust to changes in the mutation rate, detailed knowledge of this rate (which can also dynamically change in the course of the system's evolution) is not an essential training parameter. 

Stochasticity in population evolutionary dynamics can arise from both demographic and environmental fluctuations. The focus here is on demographic stochasticity arising from single-cell variability in birth, death, and mutation processes. For simplicity, spatial homogeneity and a constant environment (other than changes in drug administration) will be assumed here. To obtain a quantifiable estimate of demographic stochasticity, a stochastic physics approach was employed here (Appendix~\ref{sec:SDEderiv}) in deriving the system of stochastic differential equations describing the heterogeneous cell population time evolution. This permits an informed estimate of the extent of demographic noise from individual cells' probabilistic behavior and a subsequent quantification of an important contribution to risk in disease progression. 

The universal parameters that enter the model are the drugs that may be used with preference information (e.g. which ones are first-line vs. last-line), the appropriate discrete time intervals in which observations become available and dosing can be altered, and the likely spectrum of dose responses.  In laboratory evolution experiments~\citep{van2018experimental} this spectrum (e.g. see Fig.~\ref{fig:doseresponse}) can be obtained by allowing the bacteria to naturally evolve under different drug levels and identifying subsequent cell lines accompanied by a characterization of their dose responses  through parameter fitting of their growth curves at different drug concentrations. We note that it is not in general necessary to identify all possible and potential mutations prior to training. In particular, knowledge of all possible genetic changes is not in principle necessary: it suffices to characterize the expected relevant diversity in drug response\footnote{Such studies, however, are frequently accompanied by genotype analysis.}. The idea is to provide enough training data for the model so that new variants are unlikely to exhibit dose responses considerably different from all those already identified. During an experimental run of the trained RL model, when a newly-emergent cell type subpopulation is detected it can thus be binned without significant loss of accuracy into the state-space dimension (Section \ref{subsec:RL}) of a previously-identified variant.

\subsubsection{Observations and diagnostics}

Early, fast, and high-resolution detection of heterogeneity in drug response is crucial for identifying resistant cell subpopulations before such populations spread~\citep{koser2014whole,el2015antimicrobial,falagas2008heteroresistance}. Resistance detection may be done through phenotype-based drug susceptibility testing or by capitalizing on advances in genomic sequencing in combination with predictions of resistance from genotype~\citep{suzuki2014prediction,ruppe2017establishing}. A key aspect is the requisite ability to effectively detect low concentrations of resistant cells in an otherwise suseptible population. Single-cell analysis is showing mounting promise in this regard and is being used to resolve heterogeneity in tumors~\citep{lawson2018tumour} and in bacterial populations~\citep{rosenthal2017beyond,kelley2017new}. Direct phenotype-based detection of small resistant bacterial subpopulations in proportions as low as $10^{-6}$ of the total population was performed in~\cite{lyu2018phenotyping}. For genotype detection, single-cell sequencing~\citep{hwang2018single} can be combined with predictions of resistance from genotype for high-resolution detection of a population's resistance profile.

These advances could be applied to near-future laboratory validation of an \textit{in vitro} CelluDose controller via combination with software implementation into an automated liquid handling system in bacterial evolution experiments. Greater automation, efficiency, and standardization are nonetheless needed for these technologies to be routinely employed as part of a clinical controller apparatus, and, in part, the intention of this work is to motivate efforts in the clinical development and use of single-cell analysis for resistance profiling by demonstrating its potential utility for automated disease control.

\section{Methods}

\subsection{Stochastic modeling and simulation of cell population evolution \label{sec:biological-model}}

The evolutionary fate of an individual cell during any given time
interval depends on the timing of its cell cycle and its interaction
with and response to its environment. The model of evolutionary dynamics employed here depends on
three processes: cell birth, cell death, and inheritable changes in cell characteristics
that lead to observable changes in growth under inhibition
by an antibiotic. We
consider here drugs that suppress cell growth by inhibiting processes
essential for cell division by binding to relevant targets and thus leading
to reduced cell birth rates. The dose-response relationship can be described by a Hill-type relationship~\citep{hill1910possible,chou1976derivation}.
Here a modified Hill-type equation will be employed that includes the potential
use of multiple drugs with concentrations given by $\text{\ensuremath{\mathbf{I}}}=\left(I_{1},...,I_{m}\right)$.
The growth rate $g_{i}$ of a particular cell type $i$
as a function of the drug concentrations to which cells are exposed is thus taken to be
\begin{equation}
g_{i}\left(\mathbf{I}\right)=\beta_{i}\left(\mathbf{I}\right)-\delta_{i}=\frac{\beta_{i,0}}{1+\sum_{\alpha}\left(I_{\alpha}/\rho_{i,\alpha}\right)}-\delta_{i}\label{eq:growth}
\end{equation}
where $\beta_{i}$ is the rate of cell birth, $\beta_{i,0}$ is its birth rate in a drug-free environment, $\delta_{i}$ is
the rate of cell death, and $\rho_{i,\alpha}$ describes the extent
of resistance of cell type $i$ to drug $\alpha$. For simplicity, since drugs are assumed here to affect only birth rates, changes that confer resistance are assumed to affect only $\beta_{i}\left(\mathbf{I}\right)$ rather than $\delta_{i}$. Resources for cell growth are assumed to be limited, with the environmental carrying capacity denoted by $K$, indicating the total bacterial cell population size beyond which no further population increases take place due to resource saturation.

Evolution is an inherently stochastic process that depends on processes,
as noted above, that individual cells undergo. However, individual
(agent)-based simulations are generally very costly for larger numbers
of cells and, moreover, obscure insight into any approximate deterministic
information about the system. At very high population levels (that
can be treated as effectively infinite) a deterministic description of system dynamics is appropriate. In the dynamic growth scenario considered here, large size discrepancies can exist at any time between the various concurrent cell populations, and a single population can decay or proliferate as the simulation progresses. The extent of demographic stochasticity in such systems will thus change in the course of evolution. For efficient modeling at higher population levels it is imperative to use a simulation model whose runtime does not scale with population size, but for accurate modeling a proper estimate of stochasticity must be included. This is done here by applying
methods from stochastic physics~\citep{van1992stochastic,gardiner1986handbook,mckane2014stochastic} to derive a diffusion approximation
of a master equation describing the continuous-time evolution of the
probability distribution of the population being in some state $\left(n_{1},n_{2},...,n_{d}\right)$,
where $n_{i}$ are the (discrete) subpopulation levels for the different
cell types. This approximation relies on the assumption that the environmental
carrying capacity is large compared to individual
subpopulation levels $n_{i}$ and results in the evolution of the
system being described by a system of stochastic differential equations (SDEs)
governing the time evolution of the subpopulation levels of the different cell types given by the continuous variables $\mathbf{x}=\left(x_{1},...,x_{d}\right)$,
where $d$ is the number of distinct phenotypes that may arise.
We only consider phenotypes whose growth rate $g_{i}\left(\mathbf{I}\right)$
at nonzero drug concentrations is higher than the baseline susceptible phenotype (wildtype) due to the negative evolutionary selection experienced
by phenotypes with lower growth rates. The system evolves according
to (Appendix~\ref{sec:SDEderiv})

\begin{equation}
\frac{dx_{i}(t)}{dt}=\left(\beta_{i}\left(\mathbf{I}\right)-\delta_{i}\right)x_{i}(t)\left(1-\frac{\sum_{j=1}^{d}x_{j}(t)}{K}\right)+\sqrt{\left(\beta_{i}\left(\mathbf{I}\right)+\delta_{i}\right)x_{i}(t)\left(1-\frac{\sum_{j=1}^{d}x_{j}(t)}{K}\right)}W_{i}(t),\label{eq:SDE}
\end{equation}
where $i=1,..,d$ and the white noise $W_{i}(t)$ satisfies 

\[
\begin{cases}
\left\langle W_{i}(t)\right\rangle =0\\
\left\langle W_{i}(t)W_{i}(t')\right\rangle =\delta(t-t').
\end{cases}
\]

From a control standpoint, the advantage of using SDEs over an agent-based model is that we can capitalize on the equations-based description for trajectory estimation and use that information in feature engineering and reward assignment, as described below. Note that the stochastic noise in these equations is not put in \textit{ad hoc} but instead emerges naturally from population demographic randomness
that arises from demographic processes on the level of single cells. For more realistic dynamics at very low cell levels, when the level of a subpopulation falls below 10 cells, the number of cells is discretized by a random choice (equal probability) to round up or
down to the nearest integer cells after each stochastic simulation
step; when cell numbers fall below 1 the subpopulation level is set
to zero.

For clarity, in this paper the term \textit{decision time step} will be used to refer to the length of time between each dosing decision and thus defines the RL time step, whereas \textit{SDE simulation step} will be used to
refer to a stochastic simulation step. A single decision time step
thus contains a large number of SDE simulation steps. The evolution of Eq.~(\ref{eq:SDE}) was simulated via the Euler-Maruyama method with a step size of 0.01 over the 4-hour decision time step (unless all populations drop
to zero before the end of the decision time step is reached).

Mutations are modeled here as random events that perturb
the system of Eq.~\ref{eq:SDE} by injecting new population members into
a particular subpopulation $x_{i}$. At the start of each SDE simulation step $\tau_{\text{sim}}=0.01$, the expected number of baseline-cell type birth events is computed as $N_{\beta} = \tau_{\text{sim}}/{\beta_{bl}x_{bl}}$, where $\beta_{bl}$ is the baseline cell birth rate at the current drug concentration and $x_{bl}$ is its population size at the start of the simulation step. $N_{\beta}$ random numbers $r_i$ are then sequentially generated, and where $r_i\leq P_{\text{step,mut}}$, such that $P_{\text{step,mut}}$ is some chosen probability of mutation, a mutation occurs and an increase of 1 cell is randomly assigned with uniform probability to any of the possible (potentially-occurring) non-baseline cell types. We note that for sufficiently small $\tau_{\text{sim}}$, where the baseline population does not experience significant changes within a simulation step, this is equivalent to sequentially generating births at the appropriate time intervals and allowing mutations with probability $P_{\text{step,mut}}$. In training, $P_{\text{step,mut}}$ was set to $10^{-6}$, in keeping with typical approximate observed rates of occurrence of resistant mutant cells in bacterial populations~\citep{falagas2008heteroresistance}; subsequent testing of the policy was done on a large range of $P_{\text{step,mut}}$ values to ensure policy robustness to uncertainty in the mutation rate.

\begin{figure}[htb!]
	\begin{center}
	\begin{minipage}[t]{0.6\linewidth}
		\includegraphics[width=1\linewidth]{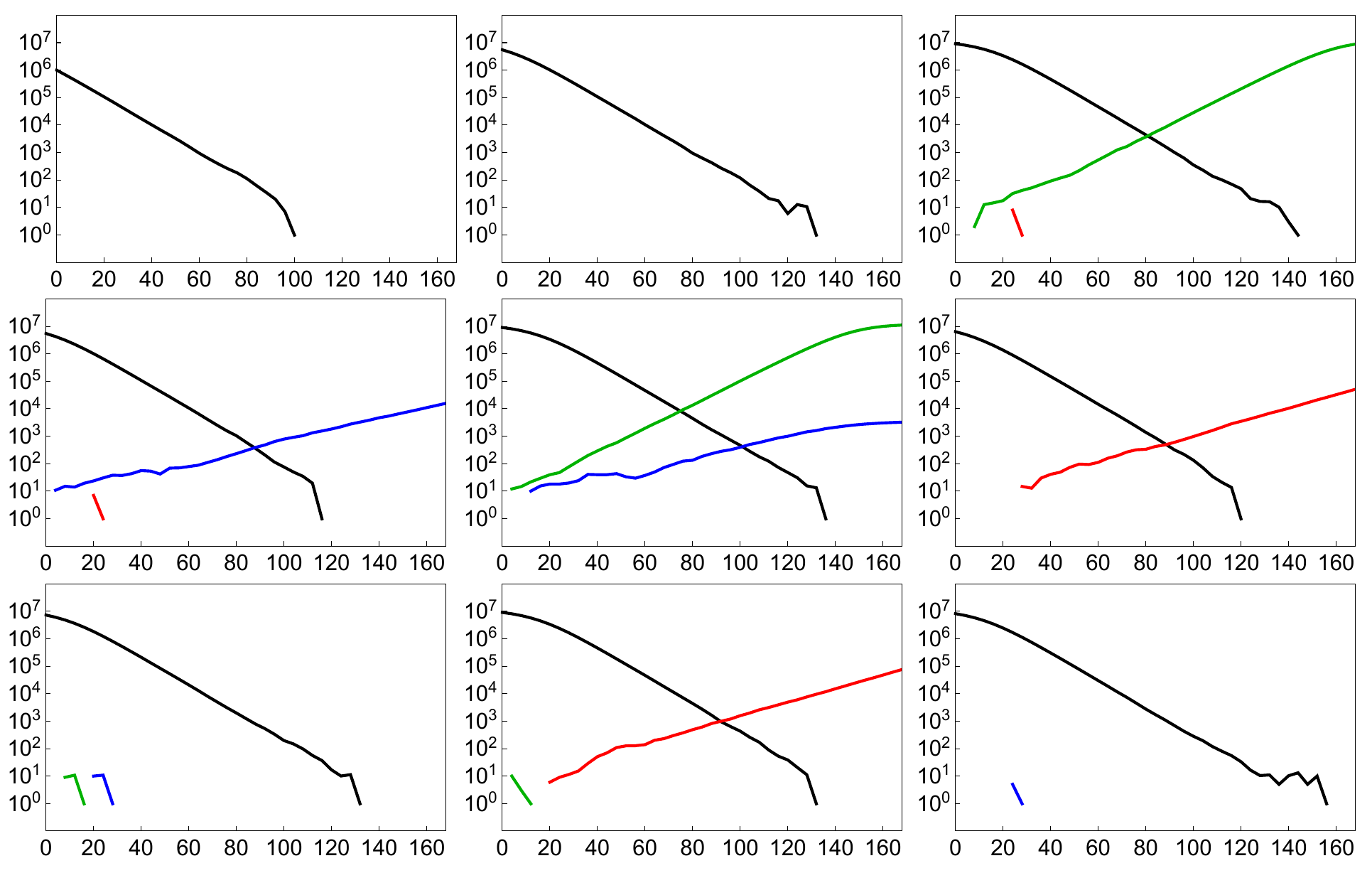}
		\captionof{figure}{\small{Sample evolutionary simulations for four phenotypes with dose-response curves shown in Fig.~\ref{fig:doseresponse}. The horizontal and vertical axes represent, respectively, time (hours) and cell concentration. The initial state in any simulation was a population comprised entirely of the (most susceptible) baseline phenotype with a population of between $10^6$-$10^7$ cells/mL (random initialization in discrete steps of $10^6$), but mutations were allowed from the first time step and could occur at any subsequent time step (4 hour intervals). The carrying capacity $K$ was set at $1.2\times 10^7$ cells/mL. The dosage was set at a constant 0.5 $\mu$g/mL for the duration of treatment, which is higher than $\rho$ for the black phenotype but lower than $\rho$ (see Eqn.~\ref{eq:growth}) for the blue, red, green phenotypes. Even when a phenotype is sufficiently strong to survive the administered dosage, random demographic fluctuations may eliminate its nascent population, as shown in several of the plots. Even when no mutations occur, variability in the initial size of the population as well as demographic randomness can lead to significantly different extinction times for the susceptible phenotype. Populations falling below 1 cell are set to zero; for plotting purposes, due to the log scale this is shown as having a population of just below 1 cell (0.98).}}
		\label{fig:simsIis1}
	\end{minipage}\hfill
	\begin{minipage}[t]{0.35\linewidth}
		\centering
		\includegraphics[width=1\linewidth]{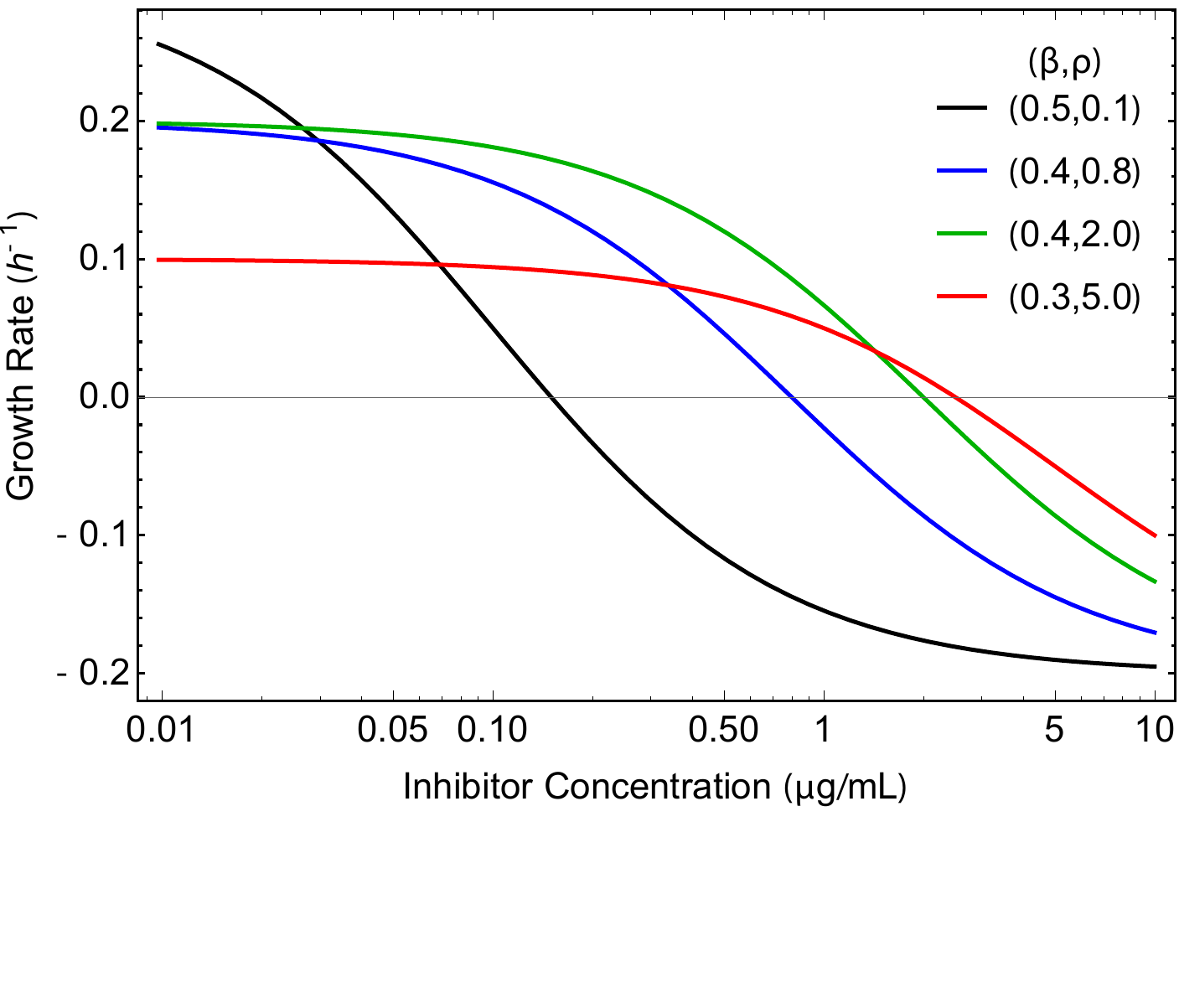}
	\captionof{figure}{\small{Dose-response curves for four sample (simulated) phenotypes, shown as growth rate as a function of drug concentration. The phenotype shown in black has the highest growth rate at zero drug inhibition. It is taken to be the baseline phenotype dominating the population at the onset of drug administration in the simulations (e.g. Fig.~\ref{fig:simsIis1}) that are used as the RL training data. The death rate was set at $\delta = 0.2$ h$^{-1}$ for all phenotypes.}}
		\label{fig:doseresponse}
	\end{minipage}
\end{center}
\label{fig:simfig}
\end{figure}

Fig.~\ref{fig:simsIis1} shows several model simulations for a constant low dose case that is above the susceptibility level of the initially-dominant phenotype but below that of three potential variants. All dose responses are shown in their respective colors in Fig.~\ref{fig:doseresponse}; the (simulated) parameters used are not specific to any particular organism or drug but fall within common concentration and growth rate ranges.

In order to permit training to identify an optimal basline policy (dosing in no-mutation episodes), during training (Section \ref{subsec:RL}) mutations were precluded from occurring during most episodes by assigning a $P_{\text{epis,mut}}=0.3$ probability that any mutations will occur during an episode (note that even when mutations are permitted to occur in an episode, they may not occur due to the low probability of mutation assigned). Notably, training with $P_{\text{epis,mut}}=0.3$ and a fixed choice of $\Phi=P_{\text{step,mut}}$ resulted in superior training performance over training with $P_{\text{epis,mut}}=1$ (mutations may occur during any episode) and a reduced probability of mutation $P_{\text{step,mut}}=0.3\Phi$ as well as over training with $P_{\text{epis,mut}}=1$ and a higher $P_{\text{step,mut}}$ for the same set of RL and neural network hyperparameters. This may be due to the combination of the comparatively higher noise, possibly reducing overfitting, together with the clear baseline signal provided by the majority of episodes.

\subsection{Learning an adaptive dosing policy\label{subsec:RL}}

As discussed in Section~\ref{subsec:RLcontrolInto}, DDPG~\citep{lillicrap2015continuous} was employed for training. The state and action spaces and reward assignment are described below, and the neural network architecture and hyperparameter choices for the
neural network and RL components are given in Appendix~\ref{sec:hyperparameters}.

In training, the maximal time for an episode was set at 7 days, with dosing decision
steps of 4 hours. Although time scales for observations will be technology-dependent
and possibly longer, this time interval was put in place in
order to demonstrate the ability of the algorithm to handle episodes
with many decision time steps, and hence a near-continuous dose adjustment,
and produce a robust control policy able to adaptively and responsively adjust to changes in the cell population composition. To put in clinical context the maximal treatment time chosen, we note that a study~\citep{singh2000short} of an intensive care unit (ICU) patient cohort found that typical ICU antibiotic courses lasted for 4-20 days, with an average length of 10 days; after 7 days of a standard antibiotic course 35\% of patients in the study were found to have developed antibiotic resistance (new resistant patterns in the original pathogen) or superinfections (newly-detected pathogenic organisms not originally present). Shorter treatment times (3 days) in that study correlated with a lower rate of antibiotic resistance (15\% at 7 days). Stopping antibiotic treatment prematurely, however, runs the risk of failing to eliminate the pathogenic bacteria, suggesting that continuous monitoring and proper dose adjustments -- as recommended here -- are essential in combatting treatment failure in severe infections.

All code (simulation and RL) was written in Python, with the use of PyTorch~\citep{paszke2017automatic} for the deep learning components.

\subsubsection{State space: model-informed feature engineering}

At each decision time step an observation of the current composition
of the cell population, i.e. types and respective concentrations $\mathbf{x}=\left(x_{1},..,x_{d}\right)$
is made. Cell types that are known to potentially arise but
are presently unobserved are assigned $x_{i}=0$. As a result of model
knowledge in the form of the SDE system (Eq.~\ref{eq:SDE}), however,
additional feature combinations of $\mathbf{x}$ can be used to supplement
and improve learning. In particular, the current growth rate of the
total cell population is highly indicative of drug effectiveness,
but it is typically difficult to measure and necessitates taking multiple
observations over some fixed time interval (e.g. 1 hour) prior to
the end of a decision time step. Instead, the instantaneous growth rate can be calculated as
\begin{equation}
g_{all}(t)=\frac{\sum_{i}\dot{x}_{i}(t)}{\sum x_{i}(t)}
\label{eq:gall}
\end{equation}
and approximated directly from observed features, $\mathbf{x}_{t}$,
at time $t$ by noting that in the deterministic limit the numerator
is simply given by 
\begin{equation}
\sum_{i}\dot{x}_{i}(t)=\left(1-\frac{\sum_{i}x_{i}}{K}\right)\times\sum_{i}g_{i}\left(\mathbf{I}\right)x_{i}(t)
\label{eq:numerator}
\end{equation}
where the $g_{i}$ are fixed and known (Eqn. \ref{eq:growth}) during any decision
time step. Hence $g_{all}$ is a function exclusively of the action
taken (doses chosen) and observations of cell concentrations $\mathbf{x}_{t}$.

Use of this feature combination, which would not be known in a blackbox
simulation, was found to be instrumental in shaping the reward signal
at non-terminal time steps and in leading to optimal policy learning,
suggesting that such incorporation of information could in general
assist in driving policy convergence when model knowledge is available
but a model-free approach is preferred due to the complexity of the
dynamics.

\subsubsection{Action space}

Training was performed with both a single drug $I_{t}$ and two drugs
$\left(I_{t}^{\alpha},I_{t}^{\beta}\right)$. The effect of the drugs on the cell populations is expressed through the dose-response as in Eq.~(\ref{eq:growth}) and the SDE system of Eq.~(\ref{eq:SDE}). Drugs were set to take a continuous range of values from zero up to some maximal value $I_{max}$, and this maximal value was used for reward scaling as shown in Section~\ref{sec:reward} rather than for placing a hard limit on the amount of drug allowed to be administered. It was set for both drugs at 8 times the highest $\rho$ value in the set $\{\rho_{i,\alpha}\}$ for all drugs $\alpha$. This very high value was chosen in order to allow enough initial dosing climb during training; in practice, conservative dosing was implemented through the reward assignment alone.

\subsubsection{Reward assignment}\label{sec:reward}

While the state space includes information on separate cell subpopulations,
the terminal (end-of-episode) reward is assigned based on only the total cell population. As noted above, a successful episode concludes in the elimination of
the targeted cell population by the maximal allowed treatment time
$T_{max}$. In training, different drugs are assigned preference through penalty weights $w_{\alpha}$, so that use of a last-line drug will incur a higher penalty than that of a first-line drug, $\alpha=1$. To penalize for higher cumulative drug dosages, the reward at the end of a successful episode is assigned as 
\begin{equation}
r_{success}=c_{end,success}\left(1-\frac{\sum_{\alpha}^{m}\left(\frac{w_{\alpha}}{w_1}\right)\sum_{n=1}^{T_{max}/\tau}I_{\alpha,n}}{c_{1}I_{max}T_{max}/\tau}\right),
\label{eq:rsuccess}
\end{equation}
where $I_{\alpha,n}$ is the dosage of drug $\alpha$ administered at the $n$th decision time step, $c_{1},\,c_{end,success}>0$ are constants, and $\tau$ is the length of each decision time step. If an episode fails (the cell population is above zero by $T_{max}$) a negative reward is assigned with an additional penalty that increases with a higher ratio of remaining cell concentrations to the initial cell population in order to guide learning:
\begin{equation}
r_{failure}=-c_{end,fail}\left[1+\log\left(1+\frac{\sum_{i}x_{i}\left(T_{max}\right)}{\sum_{i}x_{i}\left(0\right)}\right)\right],
\end{equation}
where $c_{end,fail}>0$. Since the episodes can be long, in order
to address the credit assignment problem~\citep{minsky1961steps} a guiding signal was added
at each time step in the form of potential-based reward shaping~\citep{ng1999policy}. Policy optimality was proved to be invariant under this type of reward signal, and it can thus provide a crucial learning signal in episodic tasks. It is given by 
\begin{equation}
r_{\text{shape}}(s_{t},a,s_{t+\tau})=\gamma\Phi(s_{t+\tau})-\Phi(s_{t}),\label{eq:shapereward}
\end{equation}
where $\gamma$ is the RL discount factor, $\Phi:S\rightarrow \mathbb{R}$ is the potential function, and $S$ is the state space. Since the drugs are assumed here to affect cell mechanisms responsible for cell birth, population growth provides a direct indicator of the efficacy of the drug. The potential function was therefore set here to 
\begin{equation}
\Phi=-c_{\Phi}\frac{g_{all}}{g_{max}},
\label{eq:potential}
\end{equation}
where $c_{\Phi}>0$, $g_{all}$ is given by Eq.~\ref{eq:gall}, and $g_{max}$ is the zero-drug full-population deterministic
growth rate at $t=0$ (see also Eq.~\ref{eq:numerator}):
\begin{equation}
g_{max}=\left(1-\frac{\sum_{i}x_{i}(t=0)}{K}\right)\frac{\sum_{i}^{d}x_{i}(t=0)g_{i}(I_{\alpha}=0,\:\forall\alpha)}{\sum_{i}^{d}x_{i}(t=0)}.\label{eq:gmax}
\end{equation}
Although incentives for low dosing were built into the terminal reward (Eq.~\ref{eq:rsuccess}), for guiding training toward low dosing it was necessary to provide some incentive for low dosing at each decision time step. The step reward was therefore assigned as the sum of the shaping reward and a dosage penalty: 
\begin{equation}
r_{\text{step}}=r_{\text{shape}}-\Theta\left(\sum_i x_i\right)\sum_{\alpha}^{m}w_{\alpha}\left(I_{\alpha}/I_{max}\right)^{2}\label{eq:stepreward}
\end{equation}
where $\eta_{\alpha}$ is the specific penalty for each drug, and $\Theta\left(\sum_i x_i\right)$ is a binary-valued function explained below.  For a previous implementation of step-wise action penalties with potential-based reward shaping see~\cite{petersen2018precision}. 


An issue peculiar to the problem of a population growing under limited
resources is that if insufficient dosing is applied and the population
subsequently quickly grows and reaches its carrying capacity, very
little change in growth (or population) will occur yet the state of infection will
be at its maximal worst state. However, the shaping reward (\ref{eq:shapereward})
-- which is based on changes from one decision time step
to the next -- will provide insufficient penalty for this (with the penalty arising solely from the $\gamma$ weighing). On the
other hand, as a result of the inhibitor penalty in (\ref{eq:stepreward}),
the algorithm may continue to reduce the dosing even further rather
than increasing it to escape this suboptimal policy. The result is
that the targeted cell population proliferates at maximal capacity
while dosing fluctuates around the minimal value of zero drug administration. This zero-dosage convergence problem was successfully resolved by assigning a significantly lower (20-fold) weight to the mid-episode dosage penalty if the total cell population exceeded the initial cell population by more than 1 cell for every initially present $10^4$ cells. The incorporation of the binary penalty function led to significantly improved stability and reduced sensitivity to the exact choice of $w_{\alpha}$.

\section{Training and results}\label{sec:results}

Up to four cell types with different dose responses were permitted to occur in each training scenario, with the most susceptible type initially dominating the population but with different cell populations allowed to emerge as early as the first time step. For single-drug  training the dose responses shown in Fig.~\ref{fig:doseresponse} were assumed. For combination therapy (two-drug) training, one drug was defined as the ``first-line'' drug and the second as the ``last-line'' drug via respective choices of $w_{\alpha}$. The responses of the black and blue phenotypes to both the first-line  and last-line drugs were taken to be identical to those shown in Fig.~\ref{fig:doseresponse}. The more resistant phenotypes (red and green) were assumed to be completely resistant to the first-line drug (by assigning them a very large value of $\rho$) and to have the dose response shown in Fig.~\ref{fig:doseresponse} to the last-resort drug. Successful single-drug and combination policies were identified in training, and a discussion of their main features and training progression is presented below.

In the single-drug case, in the early stages of training the dosages were increased rapidly and uniformly in all time steps; at that point dosages were too high for any failures to occur. After about 200 episodes considerable decreases in dosing began to occur (also uniformly). The rate of decrease slowed down once the dosage reached the approximate level necessary to eliminate the most resistant cell subpopulations (at about 560 episodes), with dosing still uniform over the entire course of treatment. At that point increasing specialization in dosage at different time intervals in response to different mutations began to occur, and the baseline (no-mutation) dosing continued to undergo further optimization and functional changes, alternating over the course of many episodes between increasing with time and decreasing with time as well as non-monotonic dose vs. time curves, before eventually settling into its final form of a linear monotonic mild increase after the first time step\footnote{A slightly higher dosage is seen in the first time step.}. In the early stage of training episode failures would occasionally occur; this ceased to happen later in training, at which point further improvements in learning were focused on dosing optimization. 

In dual-drug training, similar uniform increases were observed in early training for both drugs, with the first-line drug (lower penalty) initially increasing more, but with this increase reversed after about 100 episodes (recall that due to the choices of $\rho$ lower dosages of the first-line drug are needed to suppress those mutations that are not completely resistant to it compared to the dosages required to suppress mutations only affected by the last-line drug). By $\sim240$ episodes into training the first-line drug was not administered at all - but treatments were still successful since the last-line drug is capable of suppressing all cell types, albeit at a higher penalty. The last-line dosage continued to be uniformly decreased until treatment failures started occurring and was subsequently increased to the level at which it could suppress mutant populations without specialization. Decreases again began occurring at that point, and the policy also started applying the first-line drug again. Training from that point on involved specialization in response to mutation and experimentation with various amounts of the first-line and last-line drugs. When no mutations were experienced, the policy learned to administer no last-line drug at all (Fig.~\ref{fig:individualplotscombo}).

\begin{figure}
	\centering
	\begin{subfigure}{0.4\linewidth}
		\centering
		\includegraphics[width=1\linewidth]{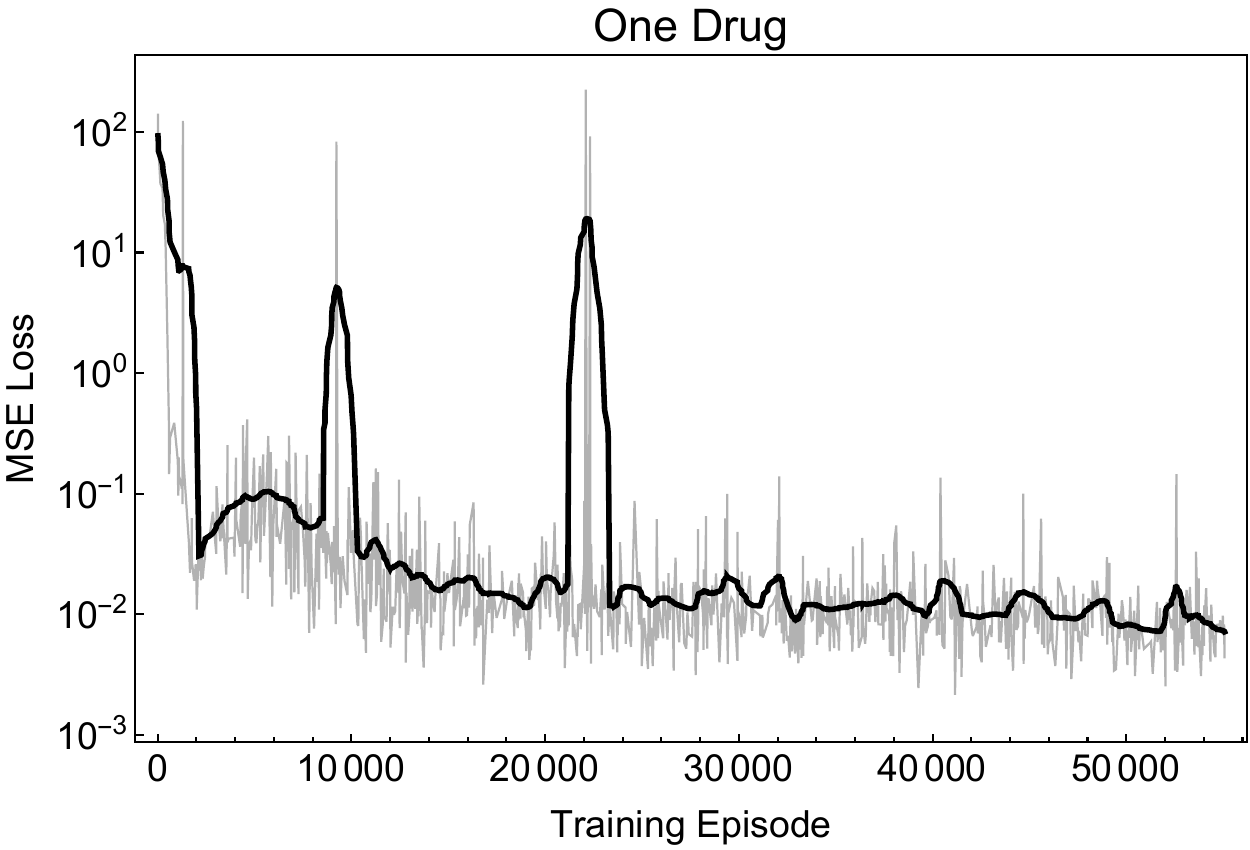}
		\caption{}
		\label{subfig:singledrugtrain}
	\end{subfigure}\hspace{0.3in}
\begin{subfigure}{0.4\linewidth}
	\centering
	\includegraphics[width=1\linewidth]{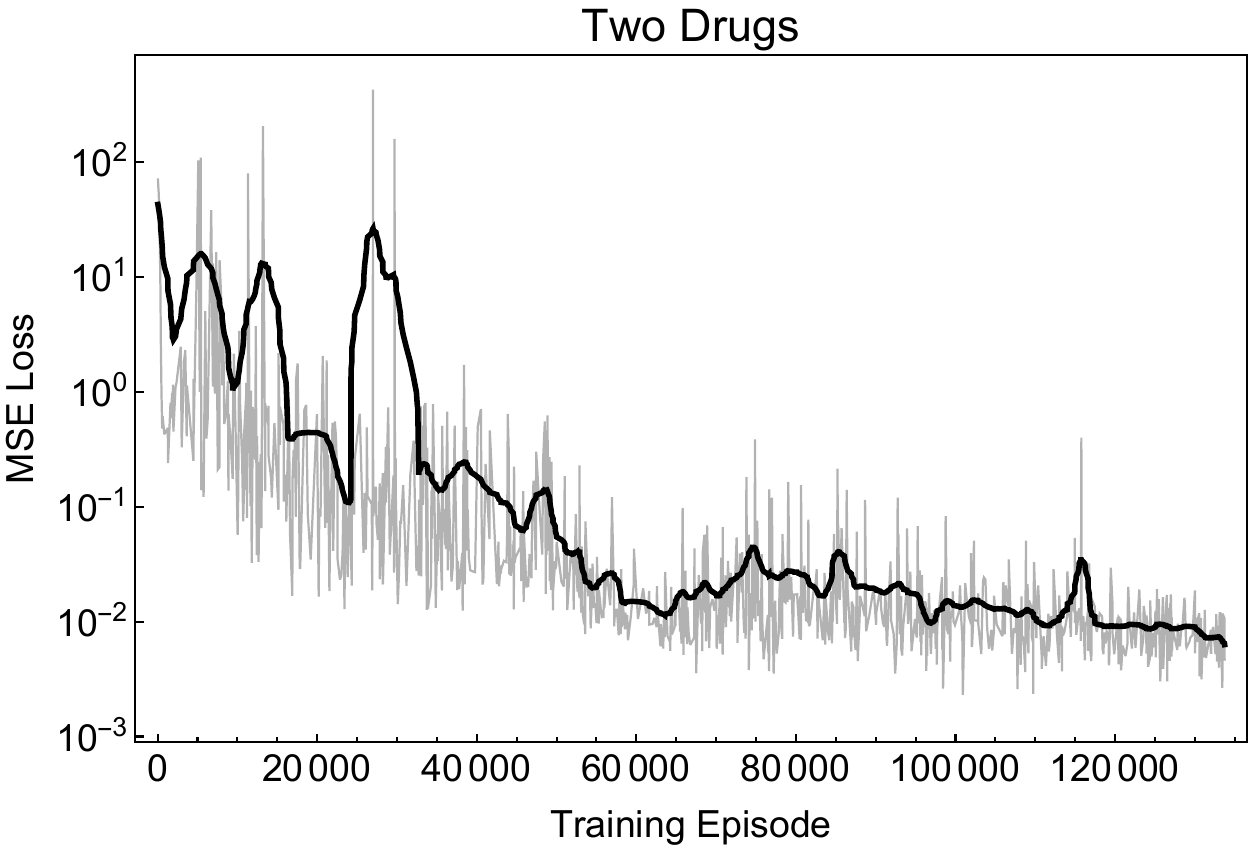}
	\caption{}
	\label{subfig:dualdrugtrain}
\end{subfigure}\\
\begin{subfigure}{0.4\linewidth}
	\centering
	\includegraphics[width=1\linewidth]{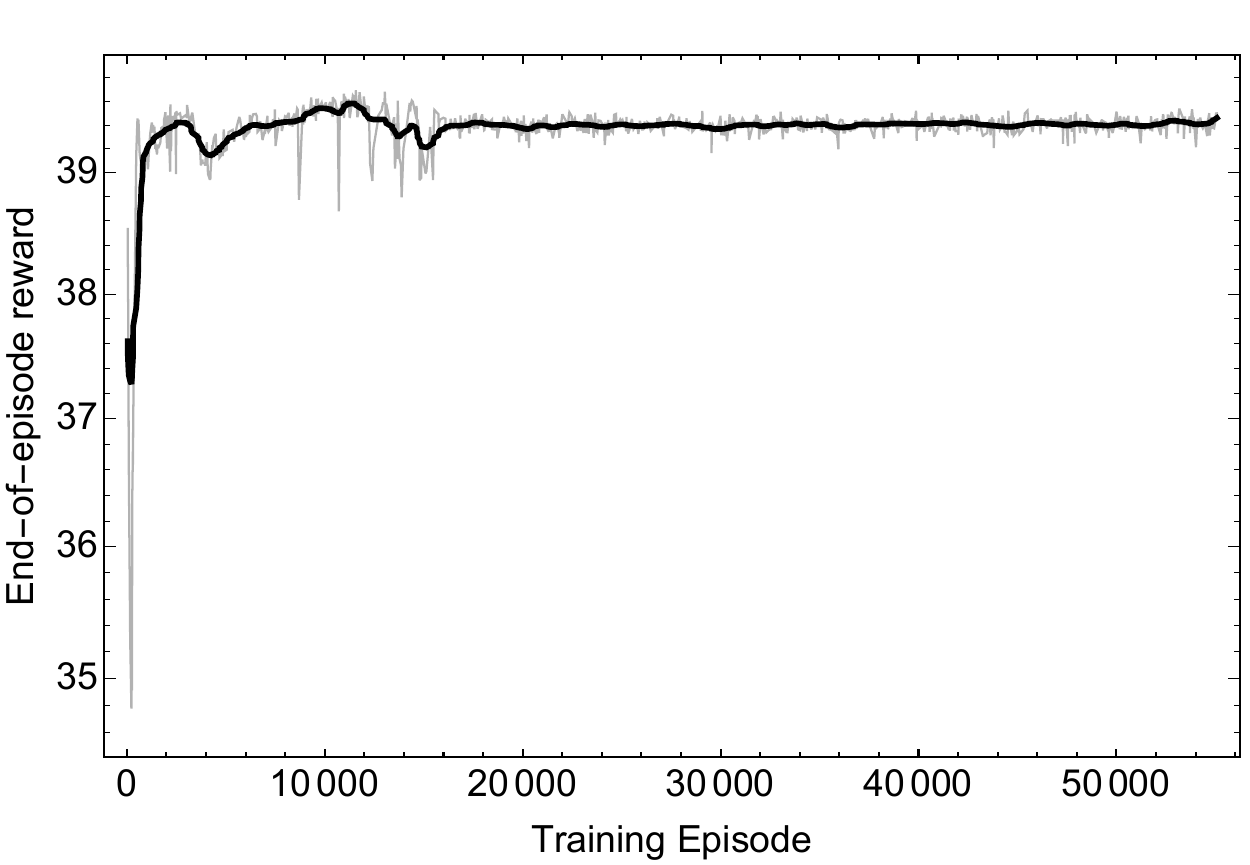}
	\caption{}
	\label{subfig:singledrugrew}
\end{subfigure}\hspace{0.3in}
\begin{subfigure}{0.4\linewidth}
	\centering
	\includegraphics[width=1\linewidth]{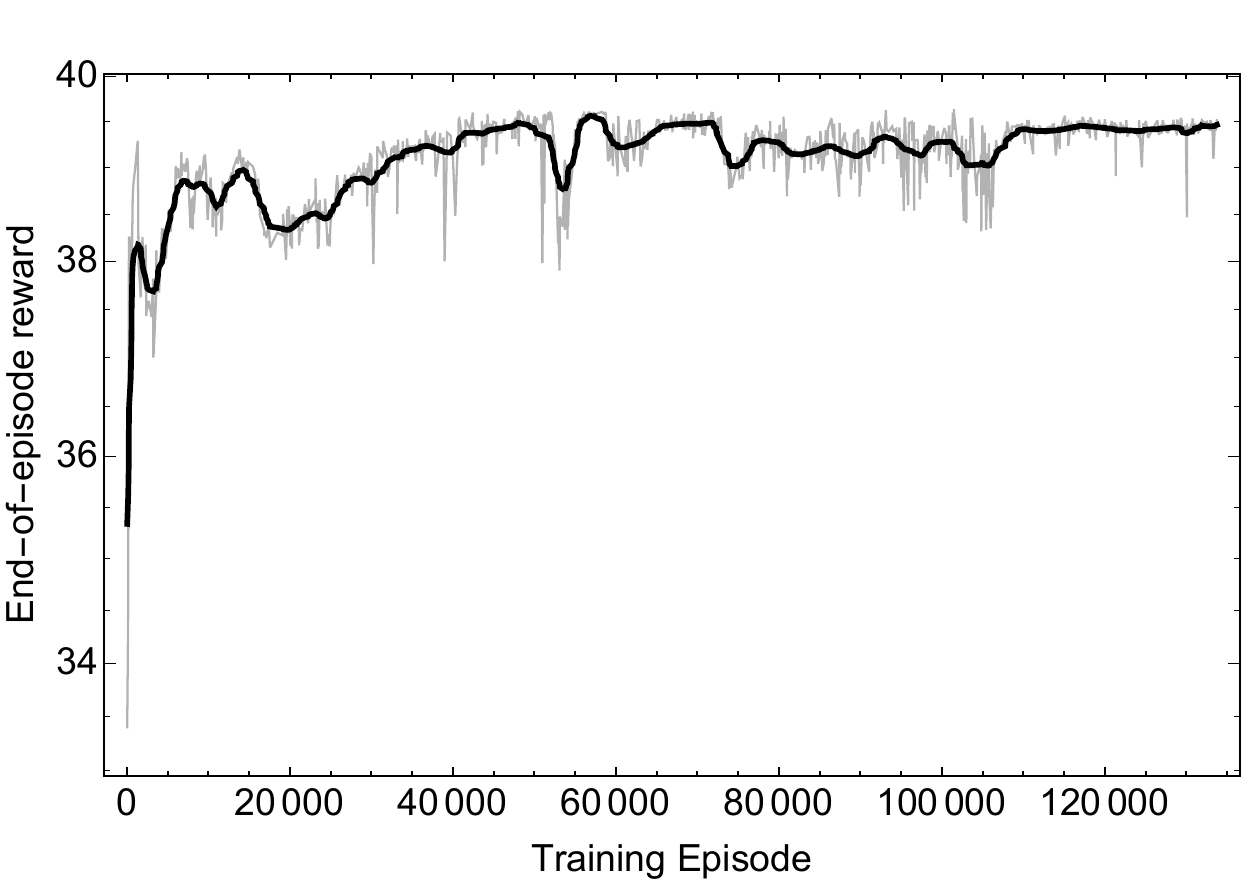}
	\caption{}
	\label{subfig:dualdrugrew}
\end{subfigure}
\caption{\small{MSE loss and end-of-episode reward when training with one drug (a and c) and two drugs (b and d). Training was generally smoother and the MSE loss flattened earlier in the one-drug case. A low-pass filter was applied to obtain the black curves. Mutation parameters for the training shown were $P_{\text{epis,mut}}=0.3$, $P_{\text{step,mut}}=10^{-6}$, and $w_1=16$, $w_2=3w_1$ for the first-line and last-line drugs respectively.}}
\label{fig:mseloss}
\end{figure}

 Training was stopped after no further reduction was observed in the MSE loss, which flattened at $\sim 3\times 10^{-3}-2\times 10^{-2}$ (with minor oscillations) in both cases, the actor policy loss had stabilized, and no further gains were observed in the end-of-episode reward signal\footnote{Note that due to the information contained in Eq.~\ref{eq:rsuccess}, measuring no further gains in this signal provides direct indication that improvements in the optimization goal of treatment success with low cumulative dosing have plateaued.}. Single-drug training converged after about 50,000 episodes or $\sim1$ million samples, taking $\sim 10$ hours on a 3.5 GhZ Intel Core i7 CPU. Dual-drug training converged after about 130,000 episodes, or $\sim 2.6$ million samples, taking a little over a day. The policy parameters were saved every 100 episodes for later reference and the resulting policies were analyzed after training. The RL training hyperparameters were identical in the single- and dual-drug training as was the depth of the neural networks. Somewhat wider hidden layers were found necessary for dual-drug training, which, as noted, also required a larger number of samples for convergence. All training hyperparameter choices are detailed in Appendix~\ref{sec:hyperparameters}.

\begin{sidewaysfigure}
	\centering
	\includegraphics[width=1\linewidth]{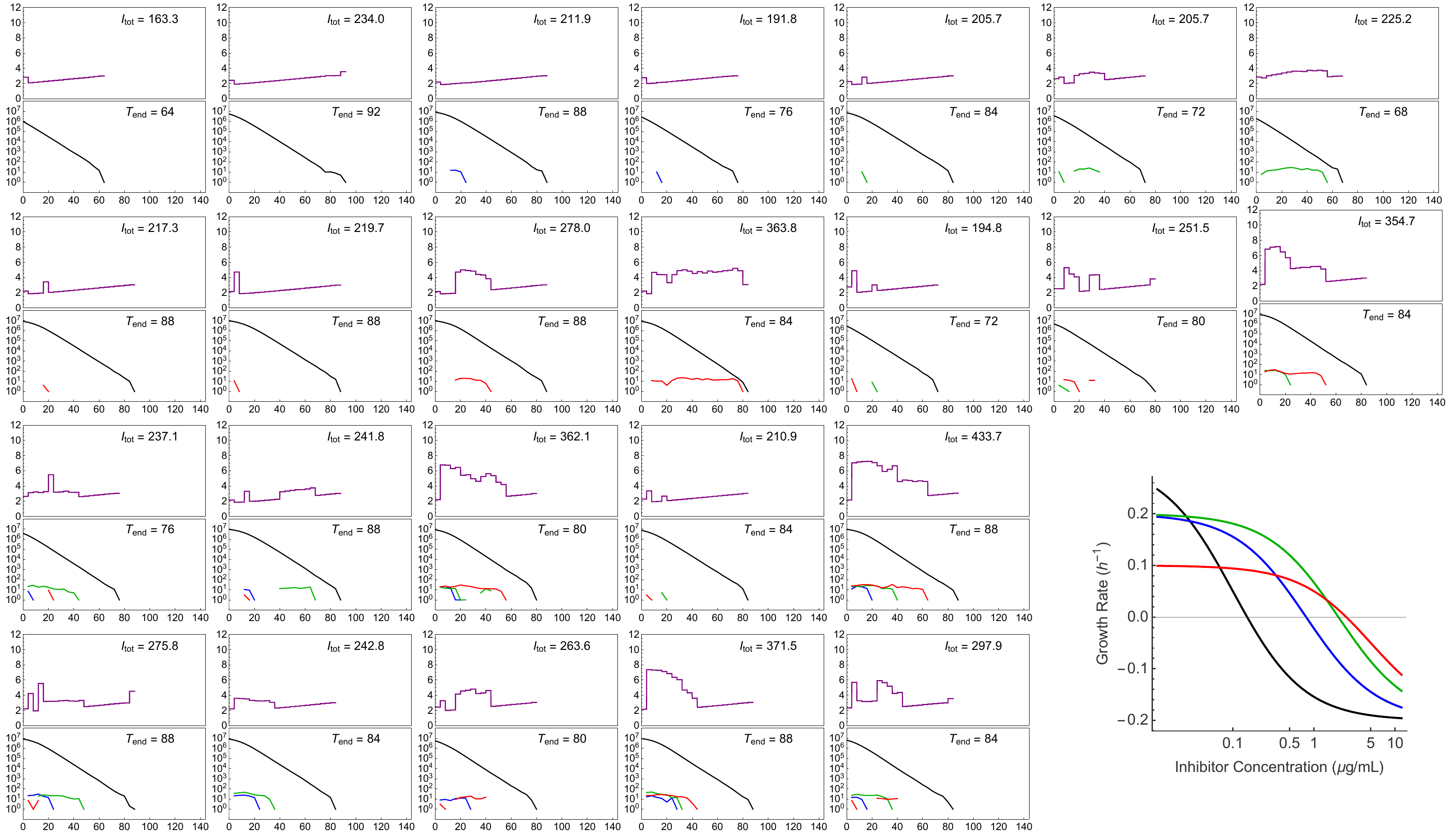}
	\caption{\small{Examples of simulation results for the combination therapy policy on a test set with mutation frequencies $P_{\text{step,mut}}$ of 1-2 orders of magnitude higher than the training value of $10^{-6}$ (all other parameters are as in Fig.~\ref{fig:simsIis1}). In each run the top plot shows drug concentration ($\mu$g/mL) as a function of time (hours) and the bottom plot shows the cell concentration as a function of time, with colors corresponding to the dose-response curves shown on the bottom right. In all cases the dosing was increased adaptively over the baseline only when necessary (note that a green population can merit higher dosing than a red one if it is observed at sufficiently high concentrations). The baseline population (black) was initialized randomly between $10^6$ and $10^7$ cells/mL in each simulation with the remaining populations initialized at zero, but resistant subpopulations were allowed to emerge as early as the first time step. The carrying capacity $K$ was set at $1.2\times 10^7$ cells/mL. Policy parameters used in these simulations were obtained after 50,000 training episodes from the training run shown in Figs.~\ref{subfig:singledrugtrain} and~\ref{subfig:singledrugrew}.}}
	\label{fig:individualplots}
\end{sidewaysfigure}
\begin{sidewaysfigure}
	\centering
	\includegraphics[width=1\linewidth]{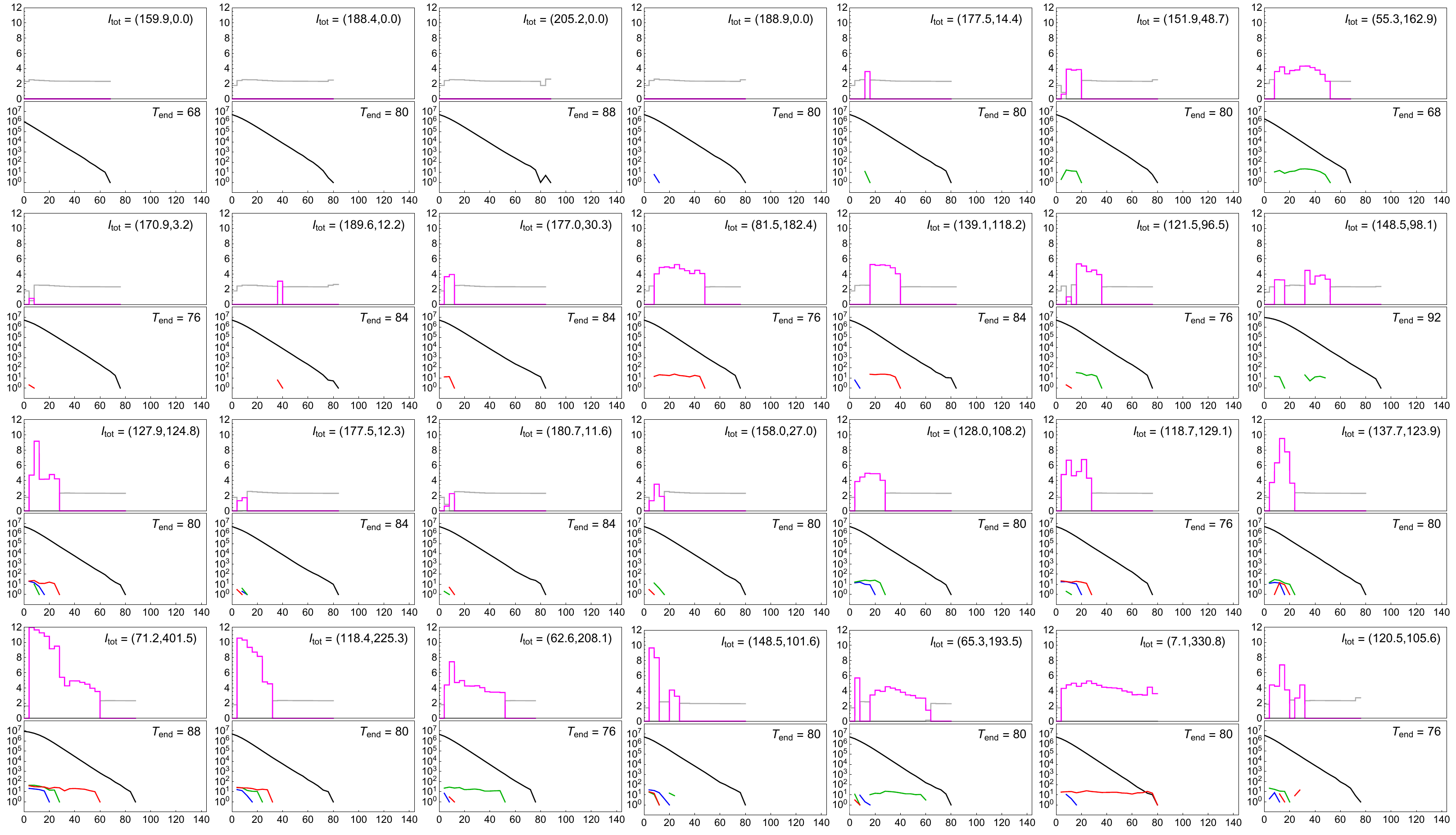}
	\caption{\small{Examples of simulation results for the combination therapy policy on a test set with mutation frequencies $P_{\text{step,mut}}$ of 1-2 orders of magnitude higher than the training value of $10^{-6}$ (all other parameters are as in Fig.~\ref{fig:simsIis1}). Both dosing adjustments of a given drug as well as drug switching took place as appropriate. The gray and magenta designate, respectively, the first-line drug and the last-line drug. In each run the top plot shows drug concentrations ($\mu$g/mL) as a function of time (hours) and the bottom plot shows the cell concentration as a function of time. $\left(I_{\text{tot,first-line}},I_{\text{tot,last-line}}\right)$ is indicated in each plot. The blue and black phenotypes exhibit the dose responses shown on the top right of Fig.~\ref{fig:individualplots} to both drugs. The green and red were assigned a $\rho$ value of 10,000 (completely resistant) to the gray drug and the same dose-response as shown on the top right of Fig.~\ref{fig:individualplots} to the magenta drug. Policy parameters used in these simulations were obtained after 130,000 training episodes from the training run shown in Figs.~\ref{subfig:dualdrugtrain} and~\ref{subfig:dualdrugrew}.}}
	\label{fig:individualplotscombo}
\end{sidewaysfigure}

\subsubsection*{Responsive adaptation to random events and system perturbations}

A hallmark of both the single-drug and combination therapy policies shown in Fig.~\ref{fig:individualplots} and Fig.~\ref{fig:individualplotscombo}, respectively, is the responsive adaptation to random population composition changes insofar as both the appearance of new cell types and stochastic fluctuations of already-present cell types. On a test set of 1000 episodes simulated randomly with training parameters a 100\% success rate was obtained in both the single-drug and dual-drug cases (all episodes recorded involved the occurrence of mutations). Multiple concurrent subpopulations are handled well by the policies learned. In single-drug therapy, after observing an emergent resistant subpopulation higher drug pulses are administered, and baseline dosing is restored following the elimination of this subpopulation (Fig.~\ref{fig:individualplots}). The dosage administered in these pulses increases with the resistance and population of the observed cell type. In combination therapy (Fig.~\ref{fig:individualplotscombo}), the policy switches to the last-line drug when the red and green phenotypes are observed and switches back to the first-line drug after these cells are eliminated. On rare occasions, when low numbers of green and red cells were present compared to the baseline (black) population, small amounts of the first-line drug were administered in tandem with the last-line drug. This may have been done in cases where the dosage of the last-line drug was enough to combat the resistant populations but not enough to sufficiently (i.e. optimally) suppress the susceptible population given its size, and compensatory increases in dosage of the first-line drug were thus given preference. In 200 simulations across above-training mutation rate values only one simulation exhibited a brief and near-negligible increase in last-line dosage when a blue cell population appeared and no first-line-resistant populations were present. Since the black and blue phenotypes are susceptible to the same extent to both the first- and last-line drugs, the sharp difference in drug administration is attributable to the differing incentives supplied in the reward assignment ($w_2=3w_1$). 

Even in the absence of mutations (the two top left plots in Fig.~\ref{fig:individualplots} and the three top left plots in Fig.~\ref{fig:individualplotscombo}) an episode can terminate at various points in time due to variations in the initial cell population size of up to an order of magnitude as well as demographic fluctuations at low populations toward the end of treatment. The dosing is seen to adjust in accordance with these variations.

\subsubsection*{Robustness to uncertainty in the mutation rate}

In a given episode, the rate and extent of resistant subpopulation emergence were controlled by $P_{\text{step,mut}}$. In reality, mutation rates can vary over time and are often unknown \textit{a priori}; the robustness of the dosing policy to this rate is therefore of critical importance to experimental and clinical applications.

To assess the robustness of the policy to variations in this parameter the success rates of the policies shown in Fig.~\ref{fig:individualplots} (one drug) and Fig.~\ref{fig:individualplotscombo} (two drugs) were tested over $P_{\text{step,mut}}$ values that exceeded the training parameter ($10^{-6}$) by several orders of magnitude (testing was done with 200 mutation-occurrence simulated episodes per value); at the high end of this range such values far exceed biologically-expected values.
No deterioration in performance was found for $P_{\text{step,mut}}$ within and substantially above the expected biological values: 100\% success was recorded  for testing of the single-drug policy for up to $P_{\text{step,mut}}=10^{-1}$ (the highest value tested) and of the dual-drug policy for up to $P_{\text{step,mut}}=10^{-3}$. The dual-drug policy exhibited degraded performance beyond a 1000-fold increase over the training value, with $47\%$ success at $P_{\text{step,mut}}=10^{-2}$ and a further increasing rate of failure beyond that point. We note that this high end of the range is in significant excess of resistant cell proportions that may be naturally expected~\citep{falagas2008heteroresistance} and testing on these ranges was performed purely for the benefit of a general robustness analysis. 

While evaluating the specific factors responsible for the high policy robustness found here is beyond the scope of this work, the randomness in the number and type of cells that was experienced during training -- through both $P_{\text{step,mut}}$ and the noise terms in Eq.~\ref{eq:SDE} -- may have contributed to the policy's ability to generalize to extended parameter ranges. This would suggest that the inclusion of some simulation stochasticity in additional model parameters, e.g. the resistance levels, could aid in training policies that are required to be highly robust against substantial parametric uncertainty in these parameters.

\subsubsection*{Learned preference for short treatment times}

\begin{figure}[ht!]
		\centering
		\includegraphics[width=0.48\linewidth]{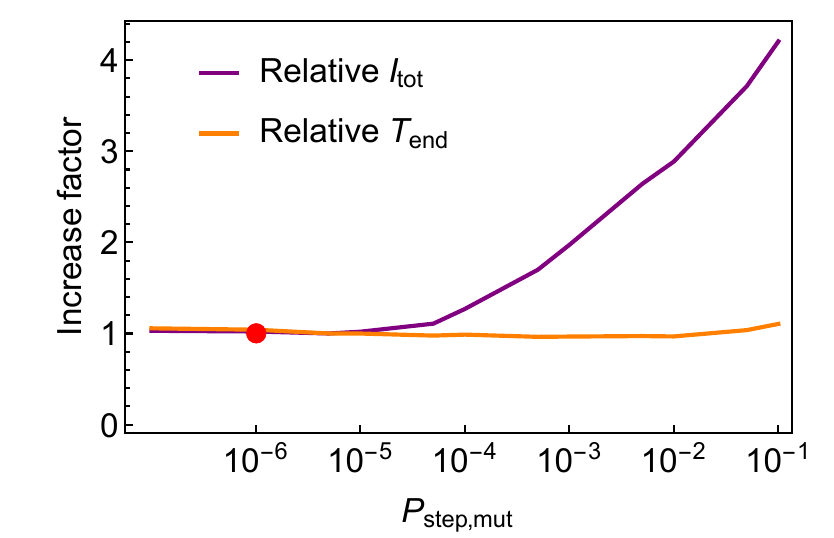}
	\caption{\small{Policy preference for early termination times can be seen by analyzing the single-drug dosing behavior (Fig.~\ref{fig:individualplots}) at different $P_{\text{step,mut}}$ parameter combinations for episodes in which mutations occurred. Data was simulated with 200 mutation-occurrence trials for each $P_{\text{step,mut}}$ value, with $P_{\text{step,mut}}$ values spaced apart by a half order of magnitude. The mean total dosage in an episode (purple) increases significantly with increases in the mutation rate, but the time of the episode (orange) remains near-constant, exhibiting a strong policy preference for maintaining short treatment times at the cost of higher baseline dosage.}}
	\label{fig:shorttimepreference}
\end{figure}

All explicit optimization incentives that were given involved the dosage administered rather than the time of treatment (Section~\ref{subsec:RL}). As previously explained, the main challenge in the dosing problem is the balancing of the higher dosages needed to suppress resistant cell populations with the need to keep toxicity low. Rewards in the setup implemented here purposely did not incorporate any direct incentives for shorter treatment times, focusing instead on dosage minimization. Although conservative dosing was a strong feature of the policies learned, these policies involved a strong preference for shorter treatment times even at the expense of higher baseline dosing, typically eliminating the cell population well in advance of the maximal allowed treatment time. This is particularly evident from a comparison of the single-drug policy response in its dosing vs. its treatment time (Fig.~\ref{fig:shorttimepreference}) to increases in the mutation probability ($P_{\text{step,mut}}$): the policy compensates for the increased resistance almost exclusively through dosage increases, while treatment time is kept nearly constant. This behavior may arise through a combination of learned preferences based on dosing penalties and mutation parameters, and in training for a real clinical scenario a ``clinician-in-the-loop'' approach would permit informed choices in this regard. The direct benefit of keeping treatment time short is that the number of resistant cells that may emerge in the course of treatment is thus reduced, avoiding further increases in dosage and/or use of last-line drugs and minimizing the chance of treatment failure due to emergent late-stage resistance.

\section{Insights for RL-based control engineering in stochastic environments}\label{sec:controlinsights}

As control systems gain in complexity, learning-based methods for control optimization may provide viable and preferred alternatives to traditional control engineering approaches. Routine use and acceptance of learning-based continuous control methods, however, will likely necessitate further improvements in training stability and convergence as well as a better understanding of how algorithmic choices affect the robustness of the learned policies. The work described in this paper suggests three particular aspects of training whose implementation could be productively generalized to other stochastic control problems employing learning-based approaches in deriving optimal policies: (1) state space augmentation via future trajectory approximation, (2) inclusion of a shaping reward that depends on the trajectory-informed dimensions of the augmented state space, and (3) generalization from model stochasticity for robustness against parameter uncertainty. Each is discussed in turn below.

In many stochastic control applications a system model in the form of component-wise equations is known or can be approximated but may have high stochasticity and/or parameter uncertainty. Dynamic control systems have time as one of their dependent variables, and the set of equations describing the system provides its time evolution, e.g., in the manner of Eq.~\ref{eq:SDE}. For notational ease, we consider here systems of ordinary stochastic differential equations, given by $\dot{x}_{i}(t)=f_{i}\left(\vec{x},\vec{\eta},t\right)$, where $\vec{x}$ is a vector of the system's components (e.g. different parts of some machinery, data flow, or biological groupings) and $\vec{\eta}$ parametrize the stochastic noise. If observations arrive as values of some or all of the components $x_{i}$, a natural state space definition consists of $\vec{x}$ or the observed subset thereof. In the case of the control problem considered in this paper, these are the populations of the different cell types at the times at which measurements are performed. In such cases, augmentation of the state space by feature combinations involving derivatives of observation space components, given by $f_{i}\left(\vec{x},\vec{\eta},t\right)$ and/or combinations thereof, essentially serves to indicate trajectory and time evolution in what would otherwise be static observations. In dynamic control situations, where the system may undergo significant changes as it evolves temporally, feeding this additional information into the function approximation mechanism used to learn the optimal control policy can mitigate the difficulties incurred in learning due to the system dynamics. Here, this was done by augmenting the observation space by an extra dimension given by $\sum_{i}\dot{x}_{i}/x_{i}=\sum_{i}f_{i}\left(\vec{x}\right)/x_{i}$ (such that stochasticity is neglected in this trajectory estimation); the inclusion of this additional state space dimension with a feature combination arising directly from the equations-based description was found to be a key driver of training convergence.

A notable feature of the state-space augmentation practiced here is that the extra state space dimension only played a role in the shaping reward (Eqs.~\ref{eq:shapereward}-\ref{eq:potential}), which provides a guiding signal in training but otherwise leaves the optimal policy invariant. This can be thought of as separating ``external'' degrees of freedom in learning -- observations and incentives that depend directly on these observations via the control optimization goal -- from ``internal'' ones that provide a guiding signal to learning through reward shaping based on state space dimensions that indicate trajectory directionality through their dependence on derivatives of the observed variables. It should be noted, however, that invariance under reward shaping involving the augmented state space does not imply that the state space augmentation itself has no effect on the optimal policy, since the extra dimension is provided as input into the actor and critic networks along with all other state space dimensions. When implementing these ideas in other environments, it may therefore be useful to assess which combinations of observations' derivatives carry information that is relevant to the optimal control goal. Here, the instantaneous growth rate, $\sum_{i}\dot{x}_{i}/x_{i}$, constitutes a relevant variable.

It is worthwhile noting that including trajectory estimation in training provides an additional level of transparency in the process of learning an optimal policy. Such added transparency can help stave off criticisms of the opacity of ``black-box'' learning and lend credence to learning-based approaches as viable control approaches that can reliably scale up to complex systems and infrastructures.

An essential aspect of establishing the reliability of a control policy operating in a stochastic environment is ensuring and testing its robustness to parameter uncertainty and noise in the system performance. Policy robustness is directly related to the generalization performance of the learning algorithm. The injection of noise into training is known to enhance neural network generalization performance, although how the extent of noise and the manner of its injection (e.g. into the training inputs, weights, etc.) precisely correlate with generalization performance is not well established. In the training described in this paper, all noise was in the training data and arose out of the natural stochasticity of the model. Interestingly, this noise was sufficient to produce policies that were not only robust to this type of noise but that were also robust to uncertainty in a key model parameter that was held fixed during training (the mutation rate). This may indicate that training on naturally highly stochastic environments could alleviate issues pertaining to parametric uncertainty and serve to increase policy robustness against this type of uncertainty.

\section{Discussion}\label{sec:discussion}

This paper presented a method for single-drug and combination therapy feedback control that was used to obtain policies capable of responsive and robust adaptation to changes in a stochastic dynamical system of evolving cell populations.  Model-free deep reinforcement learning was supplemented with model information on system trajectory estimation into feature engineering and reward assignment, which was found to significantly improve learning. Various aspects of the resulting policies were investigated in this study and may help guide future efforts in training similar systems in which some trajectory estimation is possible, there is substantial uncertainty in the probability of random perturbing events, and a tradeoff between conflicting target goals (here: low toxicity with the need to properly target resistant cell populations) exists. 

Given the demonstrated ability of the policies that were identified in training to robustly adjust to large dynamical variations on which they were not trained, an interesting question for future exploration is how accurately the underlying structure of the model -- here logistic growth dynamics -- needs to be known for policy training. Higher flexibility in this regard would enable controller development with more limited predictive knowledge of the dynamics. The resolution of feedback observations and especially their frequency is likely to play an important role in this flexibility. Here, observations were assumed to be generated at discrete time points but be reflective of the population composition at these time points. Practical implementation may necessitate accounting for a certain time delay in obtaining such observations, and future directions for the development of CelluDose include the incorporation of delayed feedback and lower-resolution observations that could lead to partial observability of the system.

As higher-resolution diagnostics, more sophisticated modeling techniques, and a recognition of the need for patient-specific care are making increasing gains in healthcare, the need for appropriate therapeutic control methods is likely to rise. The ultimate goal in precision dosing -- a feedback loop in which drug absorption and response are tightly monitored and used to determine subsequent dosing -- remains for now an elusive goal in all but a few cases~\citep{tucker2017personalized}. The method presented here is intended for eventual use in this context under consistent monitoring of the targeted cell population composition in terms of the heterogeneity in its dose responses. 

An important advantage of modeling and simulation in developing drug dosage controllers is the ability to investigate a large range of drug schedules even prior to clinical trials and to thus potentially inform the development of clinical trials. The focus of the implementation presented here is on the drug responses and trajectories of an evolving population of cells; for clinical applications, modeling would need to additionally include patient-relevant models of drug absorption and immune system response. With appropriate diagnostics, the setup described here could be implemented for the automated control of laboratory bacterial evolution experiments and establish the roles and interplay of decision time intervals, diagnostic resolution, and model uncertainty as part of the controller design. This would be a crucial step toward future clinical application and a validation of the utility of artificial intelligence-driven approaches in simulation-based biomedical controller design.


\section*{Acknowledgements}
I am grateful to Tommaso Biancalani, Roger Brockett, Finale Doshi-Velez, Thomas McDonald, Franziska Michor, and Michael Nicholson for useful discussions. This work was supported by NIH grant 5R01GM124044-02 to Eugene Shakhnovich.

\renewcommand{\theHsection}{a\arabic{section}}
\appendix
\section{Master equation to stochastic differential equations} \label{sec:SDEderiv}  

\subsection{Master equation for the population dynamics of $d$ phenotypes}

Let $A_{i}$ denote a single cell of phenotype $i$. We consider $d$
phenotypes with equal resource utilization evolving subject to a resource
capacity given by $K$. Following a simplified setup of~\cite{mckane2004stochastic}, we initially consider a spatial grid permitting up to one individual
of type $A_{i}$ per site. If empty, individual in site is denoted
by $E$. The rates of birth and death of cells of type $i$
are denoted by $\tilde{\beta}_{i}$ and $\tilde{\delta}_{i}$ respectively:
\[
\begin{alignedat}{1}\text{Birth:} & \quad A_{i}E\stackrel{\tilde{\beta}_{i}}{\rightarrow}A_{i}A_{i}\\
\text{Death:} & \quad A_{i}\stackrel{\tilde{\delta}_{i}}{\rightarrow}E
\end{alignedat}
\]
At each time step we sample the population. On a fraction $\eta$
of these events we randomly choose two individuals and allow them
to interact, but since we do not consider direct competition effects other than through shared limited resources, we ignore combinations in which both $A_i$ cells were picked and therefore restrict to picking only the combination $A_i$ and $E$. In a fraction $1-\eta$ of these events we choose only one individual randomly (if only $E$'s are drawn, the drawing is done again). The probabilities of picking the different combinations
are given by
\[
\begin{alignedat}{1}P\left(A_{i}E\right):\; & 2\eta\frac{n_{i}}{K}\left(\frac{K-\sum_{k}n_{k}}{K-1}\right)\\
P\left(A_{i}\right):\; & \left(1-\eta\right)\frac{n_{i}}{K}
\end{alignedat}
\]
where $n_{i}$ is the population of phenotype $i$. We will denote
the transition probabilities from state $\mathbf{n}=(n_{1},...,n_{d})$
to state $\mathbf{n}'$ by $T(\mathbf{n}'|\mathbf{n})$. They are
thus given by 
\[
\begin{cases}
T(n_{1},...,n_{i}+1,...,n_{d}|\mathbf{n}) & =2\eta\tilde{\beta}_{i}\frac{n_{i}}{K}\left(\frac{K-\sum_{k}n_{k}}{K-1}\right)\approx2\eta\tilde{\beta}_{i}\frac{n_{i}\left(K-\sum_{k}n_{k}\right)}{K^{2}}\\
T(n_{1},...,n_{i}-1,...,n_{d}|\mathbf{n}) & =(1-\eta)\tilde{\delta}_{i}\frac{n_{i}}{K}
\end{cases}
\]
where for simplicity we replace $K-1\rightarrow K$ as we will be
considering the large-$K$ limit for the subsequent diffusion approximation.
The master equation is given by 
\begin{equation}
\begin{aligned}\frac{dP(\mathbf{n},t)}{dt}=\sum\phantom{}_{i=1}^{d} & \left\{ T(\mathbf{n}|n_{1},...,n_{i}+1,..,n_{d})P(n_{1},...,n_{i}+1,..,n_{d},t)\right.\\
&+T(\mathbf{n}|n_{1},...,n_{i}-1,..,n_{d})P(n_{1},...,n_{i}-1,..,n_{d},t)\\
& \left.-\left[T(n_{1},...,n_{i}-1,..,n_{d}|\mathbf{n})+T(n_{1},...,n_{i}+1,..,n_{d}|\mathbf{n})\right]P(\mathbf{n},t)\right\} 
\end{aligned}
\label{eq:MasterGeneral}
\end{equation}
subject to the initial condition $P(\mathbf{n},0)=\delta_{\mathbf{n_{0},\mathbf{n}}}$
(Kronecker delta, not to be confused with the death rates $\delta_{i}$).
We must also impose boundary conditions,
\[
\begin{cases}
T(n_{1},...,n_{i}=0,...,n_{d}|n_{1},...,n_{i}=-1,...,n_{d})=0\\
T(\sum_{i}n_{i}=K|\sum_{i}n_{i}=K+1)=0.
\end{cases}
\]

\subsection{Diffusion approximation: the Fokker-Planck equation and Ito stochastic differential equations}

By assuming that the resource capacity $K$ is large compared to the
population levels $n_{i}$ (see further discussion below), we can approximate $\varphi_{i}=n_{i}/K\ll1$
as continuous variables; we denote $P(n_{1},...,n_{i}\pm1,...,n_{d},t)\equiv P_{\varphi}(\varphi_{i}\pm1)$
and define
\[
\begin{cases}
f\left(\varphi_{i}\right)\equiv\delta_{i}\varphi_{i}\\
g\left(\varphi_{i}\right)\equiv\beta_{i}\varphi_{i}\left(1-\sum_{j}^{d}\varphi_{j}\right)
\end{cases}
\]
where we have rescaled the death and birth rates as $\delta_{i}\equiv\left(1-\eta\right)\tilde{\delta}_{i}$
and $\beta_{i}\equiv\eta\tilde{\beta}_{i}$. Eqn. (\ref{eq:MasterGeneral})
can then be rewritten as 
\[
\begin{aligned}\frac{dP_{\varphi}(\vec{\varphi},t)}{dt}=\sum_{i=1}^{d} & \left\{ \left[f\left(\varphi_{i}+\frac{1}{K}\right)P_{\varphi}\left(\varphi_{1},...,\varphi_{i}+\frac{1}{K},...,\varphi_{d},t\right)-f\left(\varphi_{i}\right)P_{\varphi}(\vec{\varphi},t)\right]\right.\\
& +\left.\left[g\left(\varphi_{i}-\frac{1}{K}\right)P_{\varphi}\left(\varphi_{1},...,\varphi_{i}-\frac{1}{K},...,\varphi_{d},t\right)-g\left(\varphi_{i}\right)P_{\varphi}(\vec{\varphi},t)\right]\right\} 
\end{aligned}
\]
Taylor expanding around $\varphi_{i}$ to second order~\citep{mckane2014stochastic}
we have that
\[
h\left(\varphi_{i}\pm\frac{1}{K}\right)P_{\varphi}\left(\varphi_{1},...,\varphi_{i}\pm\frac{1}{K},...,\varphi_{d},t\right)=h\left(\varphi_{i}\right)P_{\varphi}(\vec{\varphi},t)\pm\frac{1}{K}\left.\frac{\partial\left(hP\right)}{\partial\varphi_{i}}\right|_{\vec{\varphi}}+\frac{1}{2K^{2}}\sum_{j}\left.\frac{\partial^{2}\left(hP\right)}{\partial\varphi_{j}^{2}}\right|_{\vec{\varphi}},
\]
which yields, after rescaling time as $\tau=t/K$, the Fokker-Planck
equation
\[
\frac{\partial P_{\varphi}(\vec{\varphi},t)}{\partial\tau}=\sum_{i=1}^{d}\left[\left.\frac{\partial\left(\left(f-g\right)P\right)}{\partial\varphi_{i}}\right|_{\vec{\varphi}}+\frac{1}{2K}\sum_{j}\left.\frac{\partial^{2}\left(\left(f+g\right)P\right)}{\partial\varphi_{j}^{2}}\right|_{\vec{\varphi}}\right],
\]
which corresponds to the system of Ito stochastic differential equations
\[
d\varphi_{i}=-\left(f\left(\varphi_{i}\right)-g\left(\varphi_{i}\right)\right)dt+\frac{1}{\sqrt{K}}\sum_{m}\delta_{mi}\sqrt{f\left(\varphi_{m}\right)+g\left(\varphi_{m}\right)}dW_{m}(t)
\]
where the white noise $W_{k}(t)$ satisfies 
\begin{equation}
\begin{cases}
\left\langle W_{k}(t)\right\rangle =0\\
\left\langle W_{k}(t)W_{\ell}(t')\right\rangle =\delta_{k\ell}\delta(t-t')
\end{cases}\label{eq:whitenoise-1}
\end{equation}
and $\delta_{k\ell}$ is the Kronecker delta (not to be confused with
the death rate $\delta_{i}$). We have that 
\[
\begin{aligned}f\left(\varphi_{i}\right)\pm g\left(\varphi_{i}\right) & =\left(1-\sum_{j}^{d}\varphi_{j}\right)\left[\delta_{i}\frac{\varphi_{i}}{1-\sum_{j}^{d}\varphi_{j}}-\delta_{i}\varphi_{i}+\delta_{i}\varphi_{i}\pm\beta_{i}\varphi_{i}\right]\\
& =\varphi_{i}\left(1-\sum_{j}^{d}\varphi_{j}\right)\left[\delta_{i}\left(\frac{\sum_{j}^{d}\varphi_{j}}{1-\sum_{j}^{d}\varphi_{j}}\right)+\delta_{i}\pm\beta_{i}\right].
\end{aligned}
\]
If we redfine $x_{i}\equiv K\varphi_{i}$ as the levels (concentrations)
rather than proportions of the populations, we observe that the fraction
\[
\frac{\sum_{j}^{d}\varphi_{j}}{1-\sum_{j}^{d}\varphi_{j}}=\frac{\frac{1}{K}\sum_{j}^{d}x_{j}}{1-\frac{1}{K}\sum_{j}^{d}x_{j}}
\]
approaches 1 in the large-$K$ limit, in which case we have
\[
f\left(\varphi_{i}\right)\pm g\left(\varphi_{i}\right)\rightarrow\frac{1}{K}\left(\delta_{i}\pm\beta_{i}\right)x_{i}\left(1-\frac{\sum_{j}^{d}x_{j}}{K}\right)
\]
yielding the system (\ref{eq:SDE})
\[\frac{dx_{i}(t)}{dt}=\left(\beta_{i}-\delta_{i}\right)x_{i}(t)\left(1-\frac{\sum_{j=1}^{d}x_{j}(t)}{K}\right)+\sqrt{\left(\beta_{i}+\delta_{i}\right)x_{i}(t)\left(1-\frac{\sum_{j=1}^{d}x_{j}(t)}{K}\right)}W_{i}(t)
\]
for $i=1,..,d$.  To prevent the term in the square root of the noise from
occasionally briefly dipping below zero and resulting in imaginary numbers (due to the cell subpopulations potentially dipping below zero during a simulation step before being set to zero), this term is clipped at zero. 

A few comments are due on the large-$K$ approximation above in the context of the work here. In general, as $x_j\rightarrow K$, this approximation ceases to provide a good description of the fluctuations in the system. In that regime, the large-$K$ approximation leads to deterministic dynamics and neglects the effect of population fluctuations near carrying capacity, which are expected in biological systems. This is not a concern in the scenario considered here: since $K$ is assumed here to be large, as $x_j$ approaches $K$ its dynamics become effectively deterministic and well-described by the logistic drift term alone; moreover, the treatment goal is to reduce the cell populations rather than maintain some distribution near carrying capacity, so that after the onset of treatment populations should spend little time in the vicinity of the resource capacity. As a result, any fluctuations at that point will have negligible effect on the progression of treatment. As the population decreases it moves farther away from the resource capacity and into the regime in which the diffusion term is a valid approximation of stochasticity, whose contribution to the dynamics also increases at lower population sizes due to the dependence on $x_i$. The logistic deterministic dependence, however, is relevant for modeling accuracy at the initial stages of treatment (where it can affect the growth rate) as well as for training: a finite resource capacity introduces into training the issue of low growth despite population proliferation (and hence treatment failure). This needed to be addressed with a particular reward assignment scheme, discussed in Section~\ref{subsec:RL}, and was done for potential future generalization to scenarios where the population levels may stay for longer closer to the carrying capacity due to treatment constraints. Lastly, setting a carrying capacity motivated a non-arbitrary scale in the problem for feature rescaling (Appendix~\ref{sec:hyperparameters}).

\section{Neural network architecture and hyperparamer choices}\label{sec:hyperparameters}

The actor and critic network (and respective target network) architecture
used is the same as in the original DDPG paper~\citep{lillicrap2015continuous},
with the notable difference that significant performance improvement
was obtained with the addition of an extra hidden layer (30 units for single-drug training, 45 units for dual-drug training)
in the actor network and that narrower hidden layers were
implemented in both single-drug (40 and 30 units, respectively) and dual-drug (60 and 45 units, respectively) training to avoid overfitting. All hidden layers employed rectified
linear activation functions; the output layer of the actor was set
to a hard tanh with range between 0 and a chosen maximum dose in order
to allow for the dosage to drop to exactly zero (particularly important
when multiple drugs are considered). Weights were initialized randomly
but biases were set to a positive value informed by the evolution
of the deterministic part of (\ref{eq:SDE}) (see Appendix~\ref{sec:bias}
for a detailed description). This was needed in order to prevent training from becoming trapped early on in a suboptimal policy in which the minimal control (0) was applied at every time step.

No batch normalization was used at hidden layers, but features were rescaled prior to being added to the replay buffer in the following
manner. Cell concentrations $x_{i}$ were rescaled as 
\[
x_{i}\rightarrow\frac{\log\left(x_{i}+1\right)}{\log(K)}.
\]
Since cell concentrations can vary by multiple orders of magnitude,
this was found to be necessary for smooth convergence. The growth
rate $g_{all}$ was rescaled as 
\[
g_{all}\rightarrow\frac{g_{all}-g_{min}}{g_{max}-g_{min}}
\]
where $g_{max}$ was defined in (\ref{eq:gmax}) and $g_{min}$ was
set to the negative of the death rate, $-\delta$, which is the lowest
growth rate ($\delta$ is the highest rate of decline) possible in
the system.

Adam optimization was used with learning rates of $10^{-5}$ for the
actor network and $10^{-4}$ for the critic network as learning was
found to be unstable for higher rates. Training was done with mini-batch
size of 128 and a replay buffer of size $10^{6}$. Soft updates to
the target networks were done with $\tau=0.001$ and the RL discount
factor was set at $\gamma=0.99$. Exploration was done with an Ornstein-Uhlenbeck
process with $\theta=0.15$, $\sigma=0.3$, and $\delta t=10^{-2}$.
Parameters used in the reward choice that resulted in successful training
were $c_{end,success}=40$, $w_{1}=w_{2}=1$ , $c_{1}=2$, $c_{end,fail}=20$,
$c_{\Phi}=10$. Single-drug training was done with both $\eta_1 = 16$ and $\eta_1 = 4$. Combination therapy (two drugs) training was done with $\eta_2=3\times\eta_1$ under the assumption that the drug able to effectively target the higher-resistance cells is a last-resort drug that also involves higher toxicity.

\section{Bias initialization}\label{sec:bias}

If we assume the lowest-variance policy -- constant uniform dosing -- and permit no mutations to occur, then in deterministic evolution the number of baseline-phenotype cells under constant dose $I$ starting from a population $x_0$ will be given at time $t$ by the solution of the noise-free $d=1$ system equivalent of (\ref{eq:SDE}):
\begin{equation}
x(t)=\frac{x_{0}e^{g\left(I\right)t}}{1+\frac{x_{0}}{K}\left[e^{g\left(I\right)t}-1\right]}.
\label{eq:singletypegrowth}
\end{equation}
where $g(I)=\frac{\beta}{1+\frac{I}{\rho}}-\delta$. If the population must be reduced to $x_{min}$ (here,  $<1$ cell) within a time $T_{max}$,
then the dosage must be no less than 
\[
I_{min}=\rho\left[\frac{\beta T_{max}}{\delta T_{max}-\log\left(\frac{x_{0}\left(K-x_{min}\right)}{x_{min}\left(K-x_{0}\right)}\right)}\right]
\]
for treatment to be successful. Initial cell populations were allowed
to vary by up to an order of magnitude; by taking the maximum value
allowed in this range we compute $I_{min}$ and set the bias to five
times this value or to 75\% of the maximal drug concentration - whichever is
smallest (typically $5I_{min}$). $x_{min}$ was set to just under 1 cell (0.98) as the population falling below a single cell marked the conclusion of a successful episode. Initializations
for both actor and critic networks were saved and reused in subsequent experiments.






\vskip 0.2in
\bibliography{dosing_bib}
\bibliographystyle{utphys}

\end{document}